\documentclass[a4paper,11pt]{article}
\usepackage{jheppub} 
\usepackage{cite}
\usepackage{tikz}
\usepackage{hyperref}
\usetikzlibrary{calc,decorations.pathmorphing}
\usepackage{amsmath, amssymb}
\usepackage{tikz-cd,feynmp-auto}
\usepackage{ytableau,youngtab}
\usepackage{graphicx}     
\usepackage{pifont}
\usepackage{subcaption}   
\usepackage{booktabs}
\usepackage{amsmath}
\usepackage{float}        
\usepackage{mdframed}
\usepackage{amsmath}
\usepackage{pgfplots}
\usepackage{tikz}
\usetikzlibrary{shapes.geometric, arrows.meta, positioning, fit, backgrounds}
\usepackage{amsmath}

\usetikzlibrary{shapes.geometric, arrows.meta, positioning}
\pgfplotsset{compat=1.18}
 
\definecolor{fref09}{RGB}{ 61,  9,121}
\definecolor{fref08}{RGB}{129, 26,141}
\definecolor{fref07}{RGB}{184, 55,121}
\definecolor{fref06}{RGB}{230,111, 78}
\definecolor{fref05}{RGB}{247,203, 47}

\def\be{\begin{equation}}
\def\ee{\end{equation}}
\def\bea{\begin{eqnarray}}
\def\eea{\end{eqnarray}}

\def\bg{\bar{g}}
\def\beq{\begin{eqnarray}}\def\eeq{\end{eqnarray}}
\def\ba#1\ea{\begin{align}#1\end{align}}
\def\bg#1\eg{\begin{gather}#1\end{gather}}
\def\bm#1\em{\begin{multline}#1\end{multline}}
\def\bmd#1\emd{\begin{multlined}#1\end{multlined}}

\def\D{\Delta}

\tikzstyle{Orange Dot}=[fill={rgb,255: red,241; green,143; blue,31}, draw=black, shape=circle]
\tikzstyle{Green Dot}=[fill={rgb,255: red,120; green,151; blue,95}, draw=black, shape=circle]
\tikzstyle{Cyan Dot}=[fill={rgb,255: red,0; green,159; blue,223}, draw=black, shape=circle]
\tikzstyle{White Dot}=[draw=black, shape=circle]

\tikzstyle{Dashed}=[-, dashed]
\tikzstyle{Average}=[-,dashed, draw = {rgb,255: red,120; green,151; blue,95},ultra thick]
\tikzstyle{Orange}=[-, draw = {rgb,255: red,241; green,143; blue,31},ultra thick]
\tikzstyle{Cyan}=[-, draw = {rgb,255: red,0; green,159; blue,223},ultra thick]
\tikzstyle{Arrowright}=[->]

\title{\boldmath Truncated Polyakov bootstrap }
\author[a]{Kaushik Kangshabanik,}
\author[a]{Apratim Kaviraj,}
\author[b]{Shailesh Lal}
\affiliation[a]{Department of Physics, Indian Institute of Technology - Kanpur, Kanpur 208016, India}
\affiliation[b]{Beijing Institute of Mathematical Sciences and Applications (BIMSA) Huaibei Town, Huairou District, Beijing
101408, China}

\emailAdd{ 
akaviraj@iitk.ac.in, kaushikk23@iitk.ac.in, shailesh.hri@gmail.com
}

\abstract{
  We set up a truncated numerical approach in the Polyakov bootstrap (PB) framework. We employ a gradient descent optimization to solve PB sum rules numerically, and hence solve crossing for arbitrary deformations of the generalized free field (GFF) spectrum in an iterative way starting from a perturbative solution. For unitary deformations the solutions coincide with extremal CFT spectra. But more interestingly, our findings indicate that a general crossing solution, not necessarily unitary (i.e. without positivity), can be uniquely identified by a smooth connection to GFF. The truncated Polyakov bootstrap approach is established through a number of examples in both single and mixed correlator settings.
}

\newcommand{\AK}[1]{{\color{red} \bf }}
\newcommand{\KK}[1]{{\color{blue} \bf }}
\newcommand{\SL}[1]{{\color{gray} \bf }}

\date{}

\begin{document}

\maketitle

\section{Introduction}
The conformal bootstrap is one of the most powerful Lagrangian-independent tools to obtain the OPE data of a CFT\cite{Poland:2018epd}. In the nonperturbative numerical bootstrap, CFTs are constrained by carving out the space of OPE data that are allowed by crossing and unitarity. The allowed OPE data live within bounds, and the bounds keep shrinking as one adds crossing constraints from more and more observables. A bootstrap bound is characterized by an `extremal solution' i.e. a set of CFT data that solves crossing and saturates that bound. Extremal solutions often come out to be very close to physically interesting CFT spectra, as exemplified by the 3d Ising CFT data \cite{El_Showk_2012, Kos:2016ysd}. \\ \\
CFTs in one dimension provide an important  playground to understand extremal bootstrap solutions as the latter can be linked to generalized free theories (GFFs) i.e. CFTs dual to massive free theories in AdS. Indeed the 4-point function of the lightest GFF fermion saturates the upper bound of the gap (lowest nonzero dimension) in the spectrum, while the lightest GFF boson saturates the upper bound of the OPE coefficient of the gap \cite{Mazac:2016qev, Mazac:2018mdx, mazavc2019analytic, Zan:2019fkw}. These are the simplest correlators characterized by a tower of double twist operators.   
The  GFF link was further extended in \cite{Ghosh:2025sic} through an important conjecture - any extremal solution with a fixed set of assumptions in the lower end of the spectrum asymptotes in the higher end to the GFF, i.e. the double twist spectrum. One expects that in order to see other twist families it is required to bootstrap mixed correlators \cite{kos2014bootstrapping, Ghosh:2023wjn} or higher point functions \cite{Poland:2023vpn,Poland:2023bny, Antunes:2023kyz}. 
Some of the features described above also generalize to higher dimension ($d>2$) GFFs as they maximize the twist gap i.e. the lowest dimension in any nonzero spin sector \cite{caron2021dispersive}. 
\\ \\
The positive semidefinite formulation of the 4-point crossing equation is important for obtaining bounds on OPE data, and positivity of OPE coefficients is crucial for that. But positivity is not a necessity if we want to include nonunitary CFTs in our solution space. Even for unitary theories, there are important bootstrap setups, e.g. crossing equation of a 5-point correlator or 2-point function in a boundary CFT \cite{Liendo:2012hy}, which lack positivity and hence cannot be framed as a positive semidefinite problem. Naturally one cannot talk about extremal theories in such cases. However these setups still have a GFF solution or an analogue of it (e.g. a disconnected correlator involving a cubic vertex $\langle \phi\phi\rangle\langle\phi\phi\phi\rangle$ for the 5-point problem, see \cite{Antunes:2025vvl}).  Instead of carving out islands in OPE data, one could then hope to classify the space of non-positive crossing solutions according to how they are connected to the GFF. But how does one do it systematically?\\ \\
\textit{Deformations of the free spectrum:} Consider a deformation of the 1d bosonic GFF spectrum (double twist operators) that is parametrized by some number $t$ that takes a range of values from very small to some finite number. Let us also assume that when $t$ is small we have a definite analytic solution, i.e. a set of perturbative corrections of the OPE data of the double twist operators. Now, can we construct a family of solutions $\mathcal{S}(t)$ with the lowest density of operators whose OPE data are smoothly connected to the perturbative one? Although for any  finite $t$ value there could be multiple solutions possible (note that we are allowing nonunitary theories too), we expect only one of them  to belong to $\mathcal{S}(t)$ by the smooth connection requirement. E.g. if we restrict to bootstrap constraints from a single (identical scalar) 4-point correlator, comparing with extremal theories \cite{Zan:2019fkw, Ghosh:2025sic}, $\mathcal{S}(t)$ should involve only deformations of the tower of double twist dimensions. If we add constraints from more correlators, multitwist operators will be corrected and introduced in $\mathcal{S}(t)$, again comparing with extremal theories \cite{Ghosh:2023wjn}. \\ \\ 
To find such a family of solutions we will adopt the truncated bootstrap. Truncated bootstrap, that was formulated initially in \cite{Gliozzi:2013ysa} and has seen many subsequent modifications in \cite{Gliozzi_2014, gliozzi2016truncatable, Li:2017ukc, Padayasi:2021sik,Hikami_2018, Poland:2023vpn}, is the alternative when the crossing setup cannot be framed as a positive semidefinite problem. In our approach the `truncation' will correspond to the point beyond which the operator dimensions will be that of the free double twists. Notably, in the 5-point crossing context it was shown in \cite{Poland:2023bny} that if a truncated spectrum is completed by GFF OPE data (specifically the spectrum of a disconnected correlator) beyond the truncation it also improves the power of the approach.  The reason for such improvement can be attributed to the fact that in general dimension operators at large spin asymptotes to the double twist dimension towers \cite{Fitzpatrick:2012yx,Komargodski:2012ek} - so the GFF replacement is a good representation of the full spectrum. \\ \\ 
\textit{Truncated Polyakov bootstrap:} We will set up the truncated bootstrap within the Polyakov bootstrap (PB) formulation. The 1d PB or equivalently the analytic functional bootstrap was established in \cite{Mazac:2018mdx,mazavc2019analytic} and extended for mixed and multi-point setups in \cite{Ghosh:2023wjn, Antunes:2025vvl}. They provide the 1d versions of dispersive sum rules \cite{gopakumar2021crossing, Penedones:2019tng, caron2021dispersive} -  for which the free theory spectrum becomes a manifest solution. As a consequence the dimensions of the heavier part of the spectrum (beyond truncation point) are automatically situated at the double twist values. Furthermore each PB sum rule is associated to the perturbative correction of a multi-twist operator OPE data. This feature helps in maintaining control as we move from perturbative to nonperturbative regimes.  \\ \\
Our computational strategy is to define a loss function that is essentially a sum of squares of the PB sum rules, and to minimize it by gradient descent optimization. A crucial part in this is to start from an approximate crossing solution, deform it a little in a specific way, and look for another solution close to it. We then repeat this process several times, each time starting from the previous new solution. This ``warm starting'' is very effective, as without it we have a huge multidimensional space of OPE data that gradient descent scans over, and it is easy to get stuck in a local minimum of the loss. The natural starting point, i.e.\ the first solution, is obviously the GFF, as it is a manifest solution of PB constraints.\\ \\
Our main results involve PB sum rules for single correlator as well as  multiple correlator setups. For a single correlator, any deformation to the GFF crossing solution can be obtained by deforming only the existing operators in the OPE. When additional constraints are added from other correlators, the same OPE is forced to introduce newer states. The deformations we consider include turning on an anomalous dimension for an existing GFF state, or adding a new operator in the OPE, which is allowed to have a negative OPE coefficient, hence introducing a non-unitarity.\\ \\
The truncated PB is well suited for a CFT problem when a perturbative description around a Gaussian theory is available, but we want to find the spectrum nonperturbatively. The spectrum may be non-unitary (e.g. that of a complex or disorder CFT \AK{cite}), or specific to an observable without positivity in OPE. An important example is the Ising BCFT, which is generally difficult to bootstrap due to lack of positivity. We explore it in our truncated approach in a follow-up work \cite{bcft-paper}. \\ \\
The paper is organized as follows: in section \ref{review} we briefly review the analytical functional and Polyakov bootstrap for single and mixed correlators. In section \ref{PB-GD} we introduce the truncated Polyakov bootstrap, discuss the details of the gradient descent optimization. We show the results for single correlator constraints in section \ref{single-correlator} and extend it to mixed correlators in section \ref{mixed-PB-GD}. We end with some comments on future prospects of this program in section \ref{conclusion}. Some details related to Polyakov bootstrap and  gradient descent optimization are given in the appendices.

\section{Review of Polyakov bootstrap} \label{review}

\subsection{Crossing Symmetry and Polyakov Block:}
Consider a 1d CFT correlation function of four identical external scalars of dimension $\Delta_\phi$
\begin{equation}
    \label{eq:2.2}
    \langle\phi(x_1)\phi(x_2)\phi(x_3)\phi(x_4)\rangle= \frac{1}{x_{12}^{2\Delta_{\phi}}x_{34}^{2\Delta_{\phi}}}\mathcal{G}(z)\,.
\end{equation}
Consider the operators in the OPE of $ \phi(x_i) \times \phi(x_j)$. Let us label their scaling dimensions as $\Delta$  and OPE coefficient $\lambda^{\phi\phi}_\Delta$. Then in the s-channel (12-34) the correlator $\mathcal{G}(z)$ has a conformal block decomposition
\begin{equation}
    \label{eq:2.3}
    \mathcal{G}(z)= \sum_{\Delta} a_{\Delta} G_{\Delta}(z)\,.
\end{equation}
Here $G_{\Delta}= z^{\Delta}{}_{2}F_{1}(\Delta,\Delta;2\Delta;z)$  denotes the SL$(2,\mathbb{R})$ conformal block, and $a_{\Delta}=(\lambda^{\phi\phi}_\Delta)^2$. Unless explicitly mentioned, we will refer to $a_{\Delta}$ as simply the OPE coefficient. \\ \\
$\mathcal{G}(z)$ is crossing symmetric under the exchange of $x_2\leftrightarrow x_4$ and $x_2\leftrightarrow x_3$ . This means
\begin{equation}
\label{fullcross}
\begin{split}
\mathcal{G}(z) &= \left(\frac{z}{1-z}\right)^{2\Delta_{\phi}}\mathcal{G}(1-z)\,,\\
\mathcal{G}(z)&=z^{2\Delta_{\phi}}\mathcal{G}\left(\frac{1}{z}\right)\,. \\\AK{u-channel}&\,\KK{checked}\,    
\end{split}
\end{equation}
\SL{changed typesetting \(\left(z^{2\Delta_{\phi}}\right)\rightarrow z^{2\Delta_{\phi}}\)}
Crossing symmetry of $\mathcal{G}(z)$ then translates to the equation
\begin{equation}
\label{eq:crossing}
\sum_{\Delta}a_{\Delta}F_{\Delta}(z) = 0\,,\quad F_{\Delta}(z)= G_{\Delta}(z)-\left(\frac{z}{1-z}\right)^{2\Delta_{\phi}}G_{\Delta}(1-z)\,,
\end{equation}
where \(F_{\Delta}(z)\) is called the crossing vector. \SL{minor rewording, typesetting} \\ \\
%
An important property of $\mathcal{G}(z)$ is its behavior at large $z$. This limit is the $u$-channel Regge limit. In this regime physical correlators are known to obey the following bound 
    \begin{equation}
        \label{Regge}
        \lim_{z\to\infty} z^{-\epsilon}\, \mathcal{G}(z) = 0\,, \qquad \text{for any } \epsilon>0\,.
    \end{equation}
Crossing solutions can be classified according to how they fall off at large $z$. If we impose stricter fall-offs in Regge limit then the allowed solution space shrinks, see \cite{Ghosh:2026gku}. 

\subsection{Analytic functionals and GFF solution}
It was established in a series of works
\cite{Mazac:2016qev, Mazac:2018mdx, mazavc2019analytic} that the crossing vector admits a very useful basis expansion
\begin{equation}
    \label{eq:basis}
    F_{\Delta}(z)= \sum_{n=0}^{\infty}\alpha_n(\Delta)F_{\Delta^b_n}(z)\, +\, \sum_{n=1}^{\infty}\beta_n(\Delta)\partial_\Delta F_{\Delta^b_n}(z)\,.
\end{equation}
Here the quantities $F_{\Delta^b_n}$ and $\partial_\Delta F_{\Delta^b_n}$ are the basis vectors. They are evaluated at the dimensions
\be\label{Deltan}
\Delta^b_n := 2\Delta_{\phi}+2n\,.
\ee
These dimensions correspond to the bosonic generalized free field (GFF) - see below in eqn. \eqref{gff-b-correlator}.  There is also a basis that corresponds to fermionic GFF (see \eqref{gff-f-correlator} below) - which we do not use in this paper.
In fact, there are more general choices of $\{\Delta^b_n\}$ over which the basis vectors may run \cite{Ghosh:2025sic}.

The coefficients $\alpha_n(\Delta)$ and $\beta_n(\Delta)$ in the expansion \eqref{eq:basis} are the actions of a set of linear functionals $\alpha_n$ and $\beta_n$ on $F_{\Delta}(z)$ - which are elements of the dual basis of \eqref{eq:basis}. They are called the bosonic analytic functionals.  Their actions satisfy the orthogonality properties
\begin{equation}
\begin{aligned}
    \label{eq:orthogonal}
    {\alpha}_n(\Delta^b_m)= \delta_{nm} &\quad ; \quad \partial_{\Delta} {\alpha}_n(\Delta^b_m)=-\delta_{0m}c_n\\
    {\beta}_n(\Delta^b_m)= 0 &\quad ; \quad \partial_{\Delta} {\beta}_n(\Delta^b_m)=\delta_{nm}-\delta_{0m}d_n\,,
\end{aligned}
\end{equation}
where the $c_n$ and $d_n$ are defined as
\begin{equation}
\begin{aligned}
    \label{eq:cndn}
    c_n = \frac{1}{2}\partial_n d_n
    \,, \ d_n=\frac{(\Delta_{\phi})_n^4 (4\Delta_{\phi}-1)_{2n}}{(n!)^2(2\Delta_{\phi})_n^2(4\Delta_{\phi}+2n-1)_{2n}} \,.
\end{aligned}
\end{equation}
The explicit form \cite{Mazac:2019shk, Paulos:2019gtx} of the functionals involves a $z$-integral that depends crucially on the analytic structure of the correlators and blocks, and the Regge boundedness property \eqref{Regge}. 
However, the integral representation can be somewhat cumbersome and not readily generalizable. So instead we use an alternative formulation of analytic functionals, the Polyakov bootstrap (PB). As shown in the next subsection, the main ingredient in PB is a Polyakov block which generates the functional actions $\alpha_n(\Delta)$ and $\beta_n(\Delta)$ directly - which is all one needs.

The action of the functionals on the crossing equation \eqref{eq:crossing} implies the following sum rules
\be\label{sumrules}
\sum_{\Delta}a_{\Delta}\alpha_n(\Delta)=0\,, \ \text{and}\, \ \sum_{\Delta}a_{\Delta}\beta_n(\Delta)=0\,.
\ee
We will refer to these as the PB sum rules. 
They are special as they trivialize a particular solution to crossing. Consider an OPE with $\Delta\in \{0,\Delta_n\}$ and let us fix $\Delta_0=\Delta_0^{b}=2\Delta_\phi$.
Then from the orthogonality conditions \eqref{eq:orthogonal} it can be straightforwardly shown that the sum rules have a manifest solution at $\Delta_n=\Delta^b_n$ with OPE coefficients
\be\label{gff-boson}
a_{\Delta_n}=a^b_{n}:=\frac{2\Gamma(2\Delta_\phi+2n)^2\Gamma(4\Delta_\phi+2n-1)}{\Gamma(2\Delta_\phi)^2\Gamma(4\Delta_\phi+4n-1)\Gamma(2n+1)}.
\ee
The $(a_{\Delta_n},\Delta_n)=(a^b_n,\Delta_n^b)$ is the bosonic generalized free field (GFF) spectrum - see the paragraph below. It is important to point out that $\beta_n(0)=0$, which is why $\Delta_n$ values are fixed manifestly from orthogonality. On the other hand,
the OPE coefficients are obtained from the $\alpha_n$ sum rules, since they satisfy $\alpha_n(0)=-a_{n}^b$. 

\paragraph{Bosonic and fermionic GFF}
The bosonic GFF spectrum corresponds to the 4-point function of a GFF scalar in 1d, i.e. the dual of a massive free scalar field in AdS$_2$. The correlator is given by:
\be\label{gff-b-correlator}
\mathcal{G}(z)=1+ z^{2\Delta_\phi}+\left(\frac{z}{1-z}\right)^{2\Delta_{\phi}}\,.
\ee
The three terms correspond to the Wick contraction of the GFF field.

A closely related correlator is that of a fermionic GFF field. Specifically, it is the 4-point function of a free Majorana fermion in AdS$_2$. This correlator is given by:
\be\label{gff-f-correlator}
\mathcal{G}(z)=1- z^{2\Delta_\phi}+\left(\frac{z}{1-z}\right)^{2\Delta_{\phi}}\,.
\ee
When decomposed in conformal blocks it corresponds to the dimensions $\Delta_n=\Delta_n^b+1$ with the OPE coefficients $a_{\Delta_n}$ given by the same formula \eqref{gff-boson} at the fermion $\Delta$ values. 
\be\label{fermiongff}
a_{\Delta_n}=\frac{2\Gamma(2\Delta_\phi+2n+1)^2\Gamma(4\Delta_\phi+2n)}{\Gamma(2\Delta_\phi)^2\Gamma(4\Delta_\phi+4n+1)\Gamma(2n+2)}.
\ee

\paragraph{Deformations of the GFF spectrum:} The power of the analytic functionals is more visible if we consider a deformation of the GFF spectrum. Consider a simple deformation of the GFF OPE data
\begin{align}
    \label{eq:gffdefo}
    \Delta_n &= \Delta_n^b + \gamma_n\,,\\
    a_n &= a_n^b + \delta_n\,,
\end{align}
for all $n=0,1,\ldots$. Here the anomalous dimensions $\gamma_n$ and OPE coefficient corrections $\delta_n$ are considered small. If we fix $\gamma_0$ to any nonzero value the sum rules \eqref{sumrules} can determine all the other corrections as follows:
\begin{equation}
    \label{eq:gffcorrections}
    \delta_n = \gamma_0\, c_n\,, \qquad \gamma_n = \gamma_0\, d_n\,.
\end{equation}
These follow straightforwardly from the orthogonality conditions \eqref{eq:orthogonal} of the sum rules. 
It is also possible to add a finite set of operators $\widetilde{\mathcal{O}}_0,\ldots,\widetilde{\mathcal{O}}_p$ with small OPE coefficients. These will become an additional set of inputs to the sum rule for which the corrections to GFF OPE data $\{\gamma_n,\delta_n\}$ can be computed.

Some comments are in order. The  result \eqref{eq:gffcorrections} has a nice physics interpretation - it corresponds to the deformation of a theory of a free scalar $\Phi$ in AdS$_2$  with a quartic interaction $g\,\Phi^4$. In \eqref{eq:gffcorrections} one may think of $\gamma_0$ as a placeholder for the coupling $g$. 
Interestingly, the perturbative solution coincides with an extremal bootstrap spectrum \cite{Zan:2019fkw}. The latter corresponds to what maximizes the OPE coefficient of the gap $\Delta_0=2\Delta_\phi+\gamma_0$. In the numerical optimization problem $\gamma_0$ does not need to be small, however.  Interestingly one can increase it till $\gamma_0=1$ and at that limit the extremal spectrum approaches the fermionic GFF values \eqref{fermiongff}. One cannot increase $\gamma_0$ further without violating unitarity as $\Delta_0=3$ is the maximum value of the gap \cite{Mazac:2016qev}.




\subsection{Polyakov bootstrap - single correlator}\label{ordinaryPB}
The main principle behind Polyakov bootstrap is to write the full correlator as an expansion in Polyakov blocks as follows
\be
\mathcal{G}(z)=\sum_{\Delta}a_{\Delta}P_{\Delta}(z)\,.
\ee
The defining properties of the Polyakov block $P_{\Delta}(z)$ are that it satisfies all the symmetry and analytic properties of the function $\mathcal{G}(z)$, namely:
\begin{enumerate}
    \item $P_\Delta(z)$ is a fully crossing symmetric function, similar to \eqref{fullcross}.

    \item $P_\Delta(z)$ is bounded at large $z$ (i.e.\ in the $u$-channel Regge
    limit). So it satisfies \eqref{Regge}.

    \item $P_\Delta(z)$ does not have any branch cut, unlike ordinary conformal
    blocks $G_\Delta(z)$.
\end{enumerate}
There is no unique way to define a Polyakov block, but a particularly
convenient definition is
\begin{equation}
    \label{eq:3.1.18}
    P_\Delta(z) = W_\Delta^{(s)}(z) + W_\Delta^{(t)}(z) + W_\Delta^{(u)}(z) + \lambda\, W^{(\text{con})}(z)\,.
\end{equation}
The objects $W_\Delta^{(i)}(z)$, $i=s,t,u$, are identified with tree-level
exchange Witten diagrams in AdS$_2$ with an external operator of dimension
$\Delta_\phi$ on the boundary, and $W^{(\text{con})}(z)$ is the four-point contact
diagram constructed from a $\Phi^4$ interaction in AdS$_2$ (see Appendix~\ref{app:Witten}). The three exchange diagrams map into
one another under crossing, while the contact diagram is crossing symmetric by
construction. Furthermore, they are all Regge bounded \cite{mazavc2019analytic}, and,
like ordinary correlators, they do not have branch cuts. Their expressions are
simple to write using Mellin transforms \cite{penedones2011writing, Paulos_2011}, which we discuss explicitly in
Appendix~\ref{app:Witten}.

The conformal block decomposition of the exchange and contact diagrams
$W_\Delta^{(i)}$ and $W^{(con)}$ dictates the following block decomposition of
the Polyakov block
\begin{equation}
    \label{eq:3.1.19}
    P_\Delta(z) = G_\Delta(z) - \sum_{n=0}^{\infty} \alpha_n(\Delta)\, G_{\Delta^b_n}(z) - \sum_{n=1}^{\infty} \beta_n(\Delta)\, \partial_\Delta G_{\Delta^b_n}(z)\,.
\end{equation}
It is necessary
to choose the constant $\lambda$ in the definition~\eqref{eq:3.1.18} in order
to ensure that the coefficient $\beta_0(\Delta) = 0$ for any $\Delta$. In this way it is ensured that writing the crossing symmetry of $P_{\Delta}(z)$ in terms of the crossing vectors implies the basis expansion \eqref{eq:basis}. The coefficients $\alpha_n$ and $\beta_n$ in these two definitions are equivalent and indeed match. 

\subsection{Polyakov bootstrap - mixed correlator }\label{mixed-PB}
Now consider the  correlation function of  4 distinct scalar operators $\phi_i$ with dimension $\Delta_i$ that is given by: \AK{please check again}\KK{checked}
\begin{equation}
\label{eq:mixcorr}
\langle \phi_1(x_1)\, \phi_2(x_2)\, \phi_3(x_3)\, \phi_4(x_4) \rangle
= \frac{1}{|x_{12}|^{\Delta_1+\Delta_2}\, |x_{34}|^{\Delta_3+\Delta_4}}
\left|\frac{x_{24}}{x_{14}}\right|^{\Delta_{12}}
\left|\frac{x_{14}}{x_{13}}\right|^{\Delta_{34}}
\times \mathcal{G}^{1234}(z)\,.
\end{equation}
We denote $\Delta_{ij}:=\Delta_i-\Delta_j$. 
We will refer to $\mathcal{G}^{1234}(z)$ as a mixed correlator. It has a conformal block decomposition:
\be
\mathcal{G}^{1234}(z)=\sum_{\Delta,P}a^{c}_{\Delta,P}G^c_{\Delta}(z)\,.
\ee
Here $c=(12,34)$, $(14,23)$ or $(13,24)$ denotes the OPE channel. The conformal blocks in the $(12,34)$ channel have the form: $G^{12,34}_{\Delta}(z)=z^{\Delta} \, {}_2F_1(\Delta-\Delta_{12},\Delta+\Delta_{34},2\Delta,z)$. For a mixed correlator the operators in the OPE carry two labels: dimensions $\Delta$ and parity $P=\pm 1$. So we may replace denoting an operator $\mathcal{O}_{\Delta,P}$ with its labels $(\Delta,P)$. The OPE coefficient $a^{ij,kl}_{\Delta,P}=\lambda^{ij}_{\Delta,P}\lambda^{kl}_{\Delta,P}$ where $\lambda^{ij}_{\Delta,P}$ is the 2-point OPE coefficient of $\mathcal{O}_{\Delta,P}$ in $\phi_i\times \phi_j$ OPE. 
Depending on parity we have $\lambda^{ij}_{\Delta,P}=(-1)^P\lambda^{ji}_{\Delta,P}$. \\ \\ 
The Polyakov bootstrap formulation was extended to the mixed correlator problem in \cite{Ghosh:2023wjn}. Here one takes into account the crossing symmetry under the simultaneous change $(x_2\leftrightarrow x_4, \Delta_2\leftrightarrow \Delta_4)$ or $(x_2 \leftrightarrow x_3, \Delta_2 \leftrightarrow \Delta_3)$. This motivates the following Polyakov block:
\begin{equation}
    \label{eq:mixedblock}
    \mathcal{P}^{1234}_{\Delta,P}(z) = a^{12,34}_{\Delta,P}\, \widetilde{W}^{12,34}_{\Delta,P}(z) + a^{14,23}_{\Delta,P}\, \widetilde{W}^{14,23}_{\Delta,P}(z) + a^{13,24}_{\Delta,P}\, \widetilde{W}^{13,24}_{\Delta,P}(z)
\end{equation}
where
\begin{equation}
    \label{eq:tildeW}
    \widetilde{W}^{c}_{\Delta,P}(z) = W^{c}_{\Delta,P}(z) - \sum_{\xi}\lambda^{c,\xi}_{\Delta,P}\mathcal{W}^{\text{(con)}}_{\xi}(z)\,.
\end{equation}
Here $c$ denotes the exchange channels for the tree level diagrams $W^{c}_{\Delta,P}(z)$ which exchange a bulk field dual to an operator of dimension $\Delta$ and parity $P$, with the external operators $\Delta_i$. In this setup there are at most three distinct contact diagrams $\mathcal{W}^{\text{(con)}}_{\xi}$ labeled by $\xi$. They correspond to independent 4-point interaction vertices with no. of derivatives $\le 2$ (this criterion  preserves Regge boundedness). Like before the coefficient $\lambda^{c,\xi}_{\Delta,P}$ of each contact diagram in \eqref{eq:tildeW} is fixed by elimination of sum rules - see below \eqref{eq:mixedsumrules1}.

Notice that we included the OPE coefficients into the definition of mixed correlator Polyakov block. So for every operator $\mathcal{O}_{\Delta,P}$ that appears in any channel there is a Polyakov block. When we sum over the block to write the correlator the OPE coefficient in the respective channel takes care of the contribution of $\mathcal{O}_{\Delta,P}$ in that channel. Obviously, if the operator is not present in a channel $c$ we have $a_{\Delta,P}^c=0$. \footnote{This was enforced in \cite{Ghosh:2023wjn} through the OPE orientation vector $\vec{r}_{\Delta,P}$. It is a unit normalized vector with components corresponding to channels, such that $a^c_{\Delta,P}=a_{\Delta,P}\,r_{\Delta,P}^c$. This way each operator has a single OPE coefficient $a_{\Delta,P}$ and another `quantum number' $\vec{r}_{\Delta,P}$. }\\ \\
The whole correlator can be written as 
\be\label{mixedPB1122}
\mathcal{G}^{1234}=\sum_{\Delta,P}\mathcal{P}^{1234}_{\Delta,P}(z)\,.
\ee
In this paper we will restrict to the specific case of $\Delta_3=\Delta_2$ and $\Delta_4=\Delta_1$. For this situation we have two distinct contact diagrams. The explicit forms of the Witten diagrams are reported in appendix \ref{app:Witten}. The block decomposition depends on the channel of expansion. In the $(12,21)$ expansion channel one obtains (and $c,d$ running over exchange channels (12,21), (12,12), (11,22)): 
\begin{equation}
    \label{eq:1234channel}
    \widetilde{W}^{d}_{\Delta,P} = G^c_{\Delta}(z)\,\delta_{cd} + \sum_k \alpha^{1221,d}_k(\Delta,P)\, G^c_{(12,k)}(z) + \sum_k \beta^{1221,d}_k(\Delta,P)\, \partial_\Delta G^c_{(12,k)}(z)\,.
\end{equation}
Here $(12,k)=\Delta_1+\Delta_2+k$. The decomposition in the $11$-$22$ channel is
\begin{equation}
    \label{eq:1423channel}
    \widetilde{W}^{d}_{\Delta,P} = G^c_{\Delta}(z)\,\delta_{cd} + \sum_k \alpha^{1122,c}_{11,k}(\Delta,P)\, G^c_{(11,k)}(z) + \sum_k \alpha^{1122,c}_{22,k}(\Delta,P)\, G^c_{(22,k)}(z)\,.
\end{equation}
Here $(11,k)=2\Delta_1+2k$ and $(22,k)=2\Delta_2+2k$.

Similar to ordinary PB formulation we may equate coefficients of each spurious block to zero. This gives the following sets of sum rules:
\begin{align}
    \label{eq:mixedsumrules1}
    \sum_c\sum_{\Delta,P} a^{c}_{\Delta,P}\, \alpha^{1221,c}_k(\Delta,P) &= \sum_c\sum_{\Delta,P} a^{c}_{\Delta,P}\, \beta^{1221,c}_k(\Delta,P) = 0\,,\\
    \label{eq:mixedsumrules2}
    \sum_c\sum_{\Delta,P} a^{c}_{\Delta,P}\, \alpha^{1122,c}_{11,k}(\Delta,P) &= \sum_c\sum_{\Delta,P} a^{c}_{\Delta,P}\, \alpha^{1122,c}_{22,k}(\Delta,P) = 0\,.
\end{align}
Since we have two contact diagrams, we choose the coefficients $\lambda^{c,\xi}$ in the Polyakov block by eliminating two sum rules. This is analogous to eliminating the $\beta_0$ sum rule in the identical external scalar case. For our mixed correlator we choose to eliminate $\beta^{1221}_0$ and $\beta^{1221}_1$ sum rules.

The mixed correlator sum rules shown above manifest the spectrum of a correlator of two decoupled GFF fields of dimension $\Delta_1$ and $\Delta_2$. Such a correlator exchanges identity $(\Delta=0,P=1)$ in one channel and the double twists with dimensions $\Delta_1+\Delta_2+k$ in the other channel. This property follows from their orthogonality condition which we show explicitly in appendix \ref{app:Witten} \AK{Done but verify}.
If the (decoupled) GFF correlator is slightly deformed the existing double twist data get corrected or new double twists are introduced. 
Due to their orthogonality each sum rule also captures a perturbative correction - which we indicate below:
\begin{align}\label{mixed-sumrule-property}
    &\alpha^{1221,c}_k \to (\delta a)^{12,21}_{(12,k)}\,,& &\beta^{1221,c}_k\to \gamma_{(12,k)}\,,\nonumber\\
    &\alpha^{1122,c}_{11,k} \to a^{11,22}_{(11,k)}\,,& &\alpha^{1122,c}_{22,k} \to a^{11,22}_{(22,k)}\,.
\end{align}
The first line shows the corrections to the OPE data of the double twists in $(12,21)$ channel, and the second line shows the (new) OPE coefficients of double twist operators $(11,k)$ and $(22,k)$ in the $(11,22)$ channel. 

\section{Truncated Polyakov Bootstrap by Gradient Descent}\label{PB-GD}
We next outline our method for solving a truncated bootstrap problem formulated in terms of PB sum rules. It is useful to pose the problem more generally. Consider a system of equations
\begin{equation}
\label{eq:formalsystem}
E_i\left(\boldsymbol{\theta}\right)=0\,,
\qquad
\forall\,i=1,\,2,\,\ldots,\,N\,,
\end{equation}
in variables \(\boldsymbol{\theta}\equiv \left(\theta_0,\,\theta_1,\,\theta_2,\,\ldots,\,\theta_n\right)\) valued in \(\boldsymbol{\Theta}\subseteq\mathbb{R}^{n+1}\)\,. 
Let the solution set of \eqref{eq:formalsystem} 
be denoted by
\begin{equation}
\label{eq:formalsol}
\boldsymbol{\Theta}_\ast
\equiv
\left\lbrace
\boldsymbol{\theta}\in\boldsymbol{\Theta},\,\vert\,
E_i(\boldsymbol{\theta})=0\,,\quad \forall i\,\, \right\rbrace\,.
\end{equation}
The set \(\boldsymbol{\Theta}_\ast\) may consist of a single point, multiple isolated points, or a continuous family of solutions. Next, construct scalar functions
\(\mathcal{L}\left(\boldsymbol{\theta}\right)\) 
which are positive semi-definite and the set of their global minima coincides with
\(\boldsymbol{\Theta_\ast}\) defined in \eqref{eq:formalsol}\,.
An illustrative example which will be used below is
\begin{equation}
\label{eq:formalloss}
\mathcal{L}\left(\boldsymbol{\theta}\vert \boldsymbol{\lambda}\right) = 
\frac{1}{N}
\sum_{i=1}^N 
\lambda_i 
\left[E_i\left(\boldsymbol{\theta}\right)\right] ^2\,,
\quad 
\lambda_i\in\mathbb{R}_{>0}\,\,\quad
\forall\quad i=1,\,2,\,\ldots\,,\,N\,.
\end{equation}
The problem of finding roots of \eqref{eq:formalsystem}
is therefore equivalent to finding global minima of the loss function \eqref{eq:formalloss}\,. The task is far from trivial when the search space is high-dimensional\,. In contrast, local minima are much easier to find. A variety of gradient based minimization algorithms are suited for this purpose, the simplest of which is Gradient Descent
\cite{cauchy1847methode}. 
A crude outline of the algorithm is as follows:
\begin{enumerate}
\item Initialize \(\boldsymbol{\theta}\) at some
\(\boldsymbol{\theta}^{(0)}\in\boldsymbol{\Theta}\).
\item At each iteration, update
\begin{equation}
\boldsymbol{\theta}^{(k+1)}
= \boldsymbol{\theta}^{(k)}
-\eta\,
\partial_{\boldsymbol{\theta}}
\mathcal{L}
\left(\boldsymbol{\theta}^{(k)}
\vert
\boldsymbol{\lambda}
\right),
\end{equation}
where \(\eta>0\) is the \emph{learning rate}.
\item After \(K\) updates, the algorithm terminates at
\(\boldsymbol{\theta}^{(K)}=\boldsymbol{\theta}_f\), where typically
\begin{equation}
\partial_{\boldsymbol{\theta}}\,
\mathcal{L}
\left(
\boldsymbol{\theta}_f
\middle|
\boldsymbol{\lambda}
\right)
\approx 0,.
\end{equation}
If so, one may conclude that \(\boldsymbol{\theta_f}\) is a local minimum\,.
\end{enumerate}
Gradient descent provably converges to a global minimum if \(\mathcal{L}(\boldsymbol{\theta})\) is convex in
\(\boldsymbol{\theta}\) see e.g. \cite{nesterov2013introductory}\,. This is no longer guaranteed if the system of equations \eqref{eq:formalsystem} 
has multiple solutions, in which case \(\mathcal{L}\) would be non-convex and gradient descent will converge to a local minimum, not necessarily a global one. 
However, gradient descent remains viable for solution-finding predicated on the assumption that at least one solution, 
say  \(\boldsymbol{\tilde{\theta}_\ast}\) of \eqref{eq:formalsystem} is known. In components\,,
\begin{equation}\label{eq:knownformalsol}
\boldsymbol{\tilde{\theta}_\ast} = \left(\tilde{\theta}_0,\,\tilde{\theta}_1,\,
\tilde{\theta}_2,\,\ldots,\,\tilde{\theta}_N\right)\,.
\end{equation}
We now take one of the dynamical variables, say \(\theta_0\), as externally tunable and optimize over the remaining variables using gradient descent. 
In other words, we define the reduced subset of variables \(\boldsymbol{\vartheta}\left(\theta_0\right)\)
\begin{equation}\label{eq:reduced_variables}
\boldsymbol{\vartheta}\left(\theta_0\right) = \left(\theta_1,\,\theta_2,\,\ldots,\,\theta_n\,\middle\vert \theta_0\right)\,,
\end{equation}
and optimize the loss \eqref{eq:formalloss} by evolving \(\boldsymbol{\vartheta}\left(\theta_0\right)\) under gradient descent at fixed \(\theta_0\)\,. 
In particular, we start with the solution \(\boldsymbol{\tilde{\theta}_\ast}\) of \eqref{eq:knownformalsol} and infinitesimally deform \(\theta_0\) to define variables
\(\boldsymbol{\vartheta}\left(\tilde{\theta}_0+\delta\tilde{\theta}_0\right)\). This provides a prescription for the initialization point, i.e. 
\begin{equation}
\boldsymbol{\hat{\vartheta}^{(0)}} = \left(\tilde{\theta}_1,\,
\tilde{\theta}_2,\,\ldots,\,\tilde{\theta}_n\middle\vert \tilde{\theta}_0+\delta\tilde{\theta}_0\right)\,.
\end{equation}
The point \(\boldsymbol{\hat{\vartheta}^{(0)}}\) is no longer a solution of \eqref{eq:formalsystem}, but assume that the dependence on \(\theta_0\) is smooth enough 
that the true solution \(\boldsymbol{\hat{\vartheta}_\ast}\) is still nearby and no local minima in the neighborhood have been generated by means of this deformation.
Optimizing the loss \eqref{eq:formalloss} by evolving \(\boldsymbol{\vartheta}\left(\tilde{\theta}_0+\delta\tilde{\theta}_0\right)\) under gradient descent would then expectedly yield the new location of the root. Further, we can iterate the above solution to scan over multiple instances of \(\theta_1\) and have implemented strategies for this below\,.
\subsection{Modus Operandi}
Formulating the truncated Polyakov Bootstrap in the above formalism is straightforward. A loss function corresponding to the sum rules for identical scalars, 
i.e. \eqref{sumrules} can be easily constructed in terms of the OPE data \(\{a_\Delta,\,\Delta\}\). An example in the spirit of \eqref{eq:formalloss} is
\begin{equation}
\label{F-loss}
\mathcal{F}(\{a_\Delta,\Delta\})=\sum_{n=0}^{n_\text{max}} v_n\Big[ \sum_{\Delta}a_{\Delta}\alpha_{n}(\Delta)\Big]^2\, + \, \sum_{m=1}^{m_\text{max}} w_m\Big[ \sum_{\Delta}a_{\Delta}\beta_m(\Delta)\Big]^2\,,
\end{equation}
where the  $v_n$ and $w_m$ are the counterparts of \(\lambda_i\) in \eqref{eq:formalloss}. In writing the above, we have implicitly taken \(n_{\mathrm{max}}\) and \(m_{\mathrm{max}}\) to be large enough that it is reasonable to hypothesize that every solution to crossing will lie close to a local minima in $\{a_\Delta,\Delta\}$ space. 
Our goal is to identify the minima that correspond to crossing solutions. We will take the dynamical variables \(\boldsymbol{\theta}\) to be $\{a_\Delta,\Delta\}$ and apply the gradient descent algorithm as formulated above.

In particular, consider an OPE data set $\{a_\Delta,\Delta\}_{\mathrm{min}}$ -- which we call the \emph{minimal spectrum} -- that minimizes $\mathcal{F}$ and has been identified as an (approximate) crossing solution. Now let us deform the spectrum by slightly modifying the dimension or OPE coefficient of an operator or adding one or more operator(s). This new point will no longer minimize $\mathcal{F}$. Nonetheless, if the deformation is small enough, the new minimum is expected to be in the vicinity of the starting point and can be searched for using gradient descent. 

A final detail on implementation: Formulating the sum rules themselves relies on delicate symbolic calculations which must also be automated. Mathematica is typically best suited for this, but a numerically competitive gradient descent is better formulated in a language used for numerics such as Python. For practicality we have adopted the following workaround. The sum rules are evaluated in Mathematica on a dense grid of \(\Delta\) values in the interval \(\Delta_{\mathrm{int}}\equiv \left[1,\Delta_{\mathrm{cap}}\right]\) which is capped off at some large enough \(\Delta_{\mathrm{cap}}\)\,. Subsequently, this sampled data is used along with the method of cubic Hermite interpolation -- outlined in Appendix \ref{sec:CHIP} -- to construct differentiable functions that approximate the corresponding sum rules and can be evaluated and differentiated at arbitrary \(\Delta\in\Delta_{\mathrm{int}}\). We thus implement our gradient-descent–based optimization scheme in Python/JAX \cite{jax2018github}\,. This reliance on interpolating polynomials 
constructed from grid sampling also implies that making small deformations to the minimal spectrum as above requires interpolations that can resolve such small values. 
We expect to resolve this shortly.
\subsection{Deformation around GFF}\label{GD-intro}

The advantage of using PB sum rules is that there is a natural minimal spectrum to start with: the GFF OPE data \eqref{gff-boson}, for which all $\alpha_n$ and $\beta_n$ sum rules vanish identically and $\mathcal{F}$ is manifestly at zero. This motivates the following deformation -- used here and in Section \ref{GD-no-new} --  
of the GFF data
\begin{align}\label{gff defo}
\Delta_n=\Delta_n^b+\gamma_n\,, \ a_{\Delta_n}=a_{n}^{b}+\delta_n\,,\quad n \in\{0,1,\ldots,\,N\}\,,
\end{align}
as shown in \eqref{gff-boson}\,. Our optimization problem is cast in terms of variables
\begin{equation}
    \boldsymbol{\vartheta} = \left(\delta_0, \gamma_{n>0}, \delta_{n>0}\middle\vert \gamma_0\right)\,,
\end{equation}
as described above. We use the above framework to explore how the solution evolves as the leading anomalous dimension $\gamma_0$ (equivalently $\Delta_0$) is tuned externally.
As we smoothly vary $\gamma_0$ the converged solution for one value of $\gamma_0$ is used to initialize the optimization at neighboring values, a procedure known as warm-starting.

Although we have a truncation set by $\Delta_N$, our spectrum does have operators beyond that point which are $\Delta_{n>N}=2\Delta_\phi+2n$ by default (the sum rules entering $\mathcal{F}$ will be unaffected by these dimensions if we keep $N\geq n_{\text{max}}$ or $m_{\text{max}}$).  This means any solution we find from single correlator truncated PB has the same counting (or density) of states in the OPE as that of GFF. This counting includes all the deformed and undeformed double-twist operators and excludes the operators whose OPE data have been fixed as assumptions. 
The counting increases as new operators are forced into the OPE.



\paragraph{Small deformation results:} We will first set $\gamma_0$ to a small nonzero value and try to replicate the perturbative results \eqref{gff defo} by our $\mathcal{F}$ minimization approach. 
For this we set 
$n_{\text{max}}=m_{\text{max}}=N=5$ and $\Delta_\phi=1$. Then $\mathcal{F}$ was minimized using gradient descent by warm-starting with all variables in $\boldsymbol{\vartheta}$ set to zero. The results for $\gamma_0=0.05$ and $0.1$ are summarized in Table \ref{gam0.05} and Table \ref{gam0.1}.


\begin{table}[h]
\centering
\begin{tabular}{|c|c|c|c|}
\hline
 & Grad. desc. result & Analytic (perturbative) value & deviation \% \AK{check} \\
\hline
$\gamma_1$ & $0.00886$ & $0.00833$ & $6.36$ \\
$\gamma_2$ & $0.00353$ & $0.00333$ & $6.01$ \\
$\gamma_3$ & $0.00195$ & $0.001785$ & $9.24$ \\
$\gamma_4$ & $0.00141$ & $0.00111$ & $27.0$ \\
\hline
$\delta_0$ & $-0.0902$ & $-0.1$ & $9.80$ \\
$\delta_1$ & $-0.0130$ & $-0.0123$ & $5.69$ \\
$\delta_2$ & $-0.00109$ & $-0.00102$ & $6.86$ \\
$\delta_3$ & $-8.18\times 10^{-5}$ & $-7.68\times 10^{-5}$ & $6.51$ \\
$\delta_4$ & $-5.84\times 10^{-6}$ & $-5.48\times 10^{-6}$ & $6.57$ \\
\hline
\end{tabular}
\caption{Data for $\gamma_0=0.05$ with $N=5$ ($\mathcal{F}_{\text{min}}\sim 10^{-13}$)}
\label{gam0.05}
\end{table} 
\begin{table}[h]
\centering
\begin{tabular}{|c|c|c|c|}
\hline
 & Grad. desc. result & Analytic (perturbative) value & deviation \% \AK{check} \\
\hline
$\gamma_1$ & $0.0189$ & $0.0167$ & $13.2$ \\
$\gamma_2$ & $0.00761$ & $0.00667$ & $14.1$ \\
$\gamma_3$ & $0.00404$ & $0.00357$ & $13.2$ \\
$\gamma_4$ & $0.00187$ & $0.00222$ & $15.8$ \\
\hline
$\delta_0$ & $-0.163$ & $-0.2$ & $18.5$ \\
$\delta_1$ & $-0.0275$ & $-0.0246$ & $11.8$ \\
$\delta_2$ & $-0.002326$ & $-0.00205$ & $13.5$ \\
$\delta_3$ & $-1.75\times 10^{-4}$ & $-1.53\times 10^{-4}$ & $14.4$ \\
$\delta_4$ & $-1.240\times 10^{-5}$ & $-1.09\times 10^{-5}$ & $13.8$ \\
\hline
\end{tabular}
\caption{Data for $\gamma_0=0.1$ with $N=5$ ($\mathcal{F}_{\text{min}}\sim 10^{-12}$)}
\label{gam0.1}
\end{table}

The general increase in deviation \% of the variables as $\gamma_0$ goes from $0.05$ to $0.1$ is expected since we are moving away from perturbative regime. It may seem that if we keep setting $\gamma_0$ to smaller values all the variables will approach the exact perturbative result. However one has to be careful since we are working with interpolated functionals that cannot be trusted beyond a certain precision. Recall that perturbatively $\gamma_n, \delta_n \propto \gamma_0$, so for very small $\gamma_0$ we should be working with the interpolations that can adequately resolve such small values. This also makes it challenging to obtain the higher OPE data, i.e. $\gamma_n, \delta_n$ for arbitrarily large $n$. This is because the corrections roughly fall off as $1/n^2$. This difficulty persists for any small deformation in the lower end of the spectrum - because the higher $n$ corrections come mainly from higher sum rules which are less dominated by the lower spectrum, see eqn. \eqref{alphabeta-small-gamma} below.

\section{Truncated Polyakov Bootstrap for a single correlator}\label{single-correlator}
We now extend the analysis above  to the case where the deformations are not necessarily small. We restrict to single correlator constraints in this section. Two classes of deformations are considered below: one that does not involve an extra state, and one that does. In both cases we track the flow of corrected GFF states.  


\subsection{Free boson deformation with no extra state}\label{GD-no-new}
This is simply an extension of the analysis of section \ref{GD-intro} into a more nonperturbative regime. We will increase $\gamma_0$ from $0.1$ all the way to $1$ in small steps.  The point $\gamma_0=1$ corresponds to the maximum gap in a unitary theory. We will see that the resulting spectrum from the truncated PB matches with the solutions that maximize the OPE coefficient of the gap $\Delta_0=2\Delta_\phi+\gamma_0$ to a high degree of accuracy. 

\paragraph{Fixing the weights/normalization:} Let us give a proper prescription for fixing the weights $v_n$ and $w_m$ in $\mathcal{F}$ as in \eqref{F-loss}. To understand their role better consider the result from $\gamma_0=0.1$ in Table \ref{gam0.1}. If we use this data but with $\gamma_0=0.2$ (say) then the first few individual sum rules are:
\begin{align}\label{alphabeta-small-gamma}
&\alpha_0\approx 0.119, & \, &
    \beta_1\approx -0.0311 \nonumber\\ &\alpha_1\approx 0.0394, & \, &
    \beta_2 \approx  -0.00240,\nonumber\\
    &\alpha_2 \approx 0.00310,  & \,  &
    \beta_3\approx -0.000177,\\
    &\alpha_3\approx 0.000231, & \,  &
    \beta_4\approx -0.0000148,\nonumber\\
    &\alpha_4\approx 0.0000414. & \,  &\nonumber
\end{align}
Both $\alpha_n$ and $\beta_n$ get smaller as $n$ increases.  If a certain sum rule is initialized with a very small value then it becomes hard to minimize it further. Therefore a reasonable choice of $v_n$ and $w_m$ is setting them proportional to $(\alpha_n)^{-1}$ and $(\beta_m)^{-1}$ respectively. This way all the sum rules become comparably important under the minimization of $\mathcal{F}$.\footnote{In principle we may also set the weights as finite numbers, e.g. $v_n=w_m=1$. But then the  $\mathcal{F}$ needs to be minimized to smaller values to accurately capture higher sum rules. The precision of interpolation of sum rules from the $\Delta$ grid also has to be higher. Naturally this way is harder to control} We can vary the constants of proportionality - which gives a way to estimate the error in the results, see section \ref{error-section}.
%


\paragraph{Warm starting:}
We, once again, assume the spectrum from \eqref{gff defo}. We increase the input deformation $\gamma_0$ from $0$ to $1$ in discrete steps:
from $0$ to $0.9$ in steps of $0.05$, and from $0.9$ to $1$ in steps of $0.01$. This distinction is simply from the observation that the changes of results in between steps were more prominent close to $\gamma_0=1$.
Consider the $i$-th step for which let us denote the input $\gamma_0$ as $\gamma_0^{(i)}$. For the other variables,
let us call their initial values as $\gamma_n^{(i),0},\delta_n^{(i),0}$ and the final values after minimization $\gamma_n^{(i),\text{min}},\delta_n^{(i),\text{min}}$. The $i$-th initial values are set as
\begin{align}\label{initial}
&\gamma_n^{(i),0} = \begin{cases} \gamma_n^{(i-1),\text{min}}  \hspace{1.5cm} \text{if } \gamma_n^{(i-1),\text{min}} \geq \gamma^{\text{pert}}_n(\gamma_0^{(i)}) \\ \gamma^{\text{pert}}_n(\gamma_0^{(i)})  \hspace{2cm} \text{otherwise\,,} \end{cases}\\
&\delta_n^{(i),0} = \begin{cases} \delta_n^{(i-1),\text{min}}  \hspace{1.5cm} \text{if } \delta_n^{(i-1),\text{min}} \geq \delta^{\text{pert}}_n(\gamma_0^{(i)}) \\ \delta^{\text{pert}}_n(\gamma_0^{(i)})  \hspace{2cm} \text{otherwise\,.} \end{cases}\label{initial-2}\
\end{align}
Here $\gamma^{\text{pert}}_n$ and $\delta^{\text{pert}}_n$ are the same perturbative values as in \eqref{eq:gffcorrections} which are functions of $\gamma_0$. The above choice of initialization is motivated from the fact that we want to set the initial $i$-th OPE data as close as possible to the correct values for $(i-1)$-th step.  
For a small deformation (i.e. $\gamma_0$) higher OPE corrections are harder to obtain accurately (see discussion above section \ref{GD-no-new}) - and they settle at values lower than the analytic ones. Clearly analytic values are closer to the correct one in this $\gamma_0$ range. As we increase $\gamma_0$ more of the data $\gamma_n^{(i),\text{min}},\delta_n^{(i),\text{min}}$ exceed the analytic values - and these values automatically become a better choice of initialization.\\ \\ 
We can, however, be flexible with the warm starting scheme \eqref{initial}-\eqref{initial-2}. E.g. instead of the analytic values, we may set some of the higher corrections to zero or small random numbers. We took into account such variations to estimate the error in section \ref{error-section}. It is observed after minimization that the lower OPE corrections are stable under changing the initial values of the higher ones. But if we change the initial values of too many of them we end up in a different local minima of $\mathcal{F}$ - which is deemed unphysical for being unstable under addition of more sum rules or initial values. This shows why the initialization is an important component of the truncated PB.


\begin{figure}
    \centering
\includegraphics[scale=0.5]{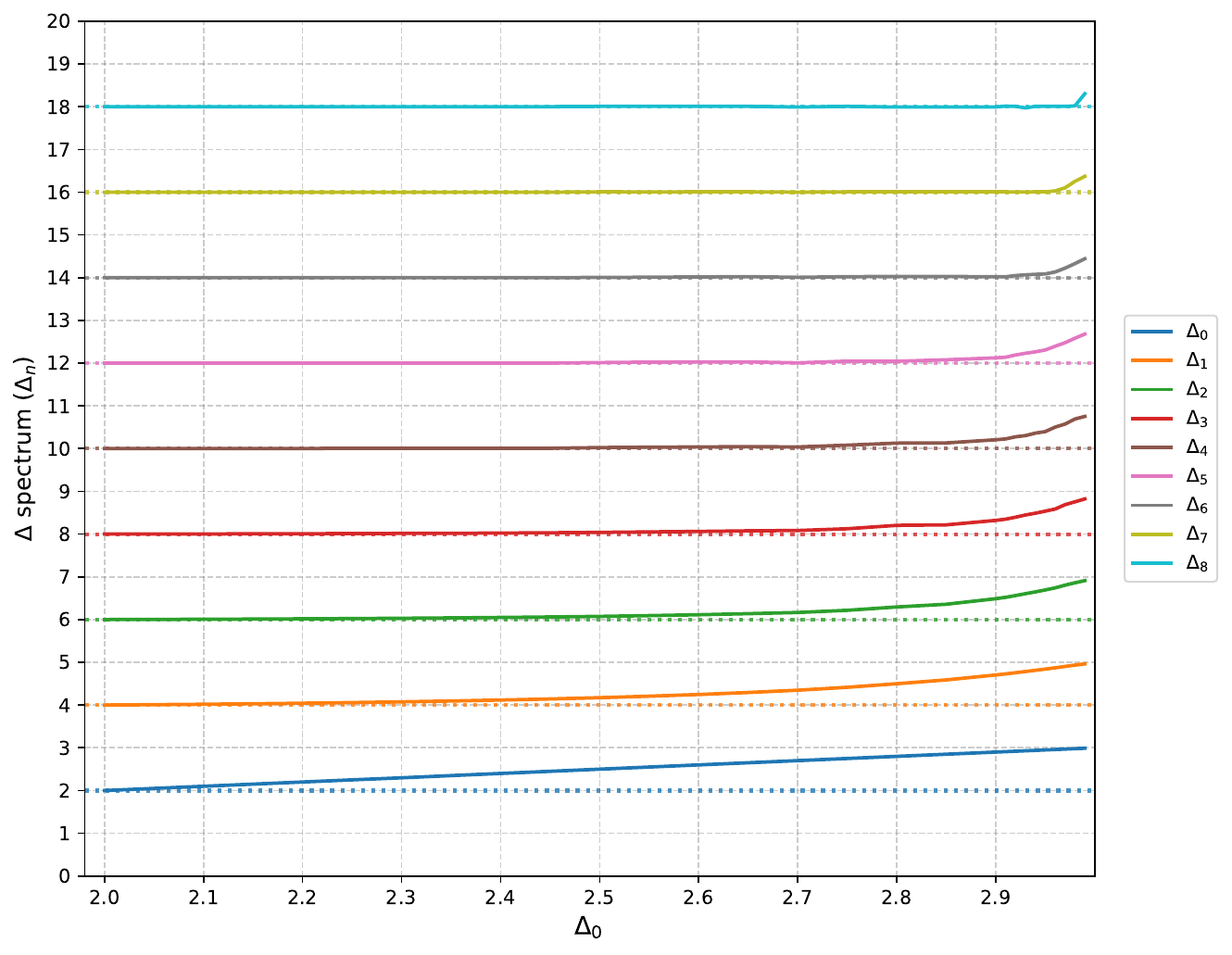}
    \caption{Flow of the minimal spectrum.}
\label{fig:spectrum}
\end{figure}
\paragraph{Flow of the minimal spectrum:} With the above rules in place we considered a loss function with 17 sum rules  and 9 operators. The results of minimization for each step are shown below. In Fig. \ref{fig:spectrum} we show the  minimal spectrum dimensions. 
The qualitative features of the plot is same as the family of extremal spectra that maximize the OPE coefficients of the operator $\Delta_0$. Towards the right the dimensions $\Delta_n$ approach $3+2n$ i.e. the GFF fermion values. The approach to the fermionic values becomes sharper with increasing $n$ - this is not surprising as there is no single Regge-bounded contact diagram deformation around GFF fermions\cite{kausik2021charging}. \AK{Move this text?}

\begin{figure}
    \centering
\includegraphics[scale=0.5]{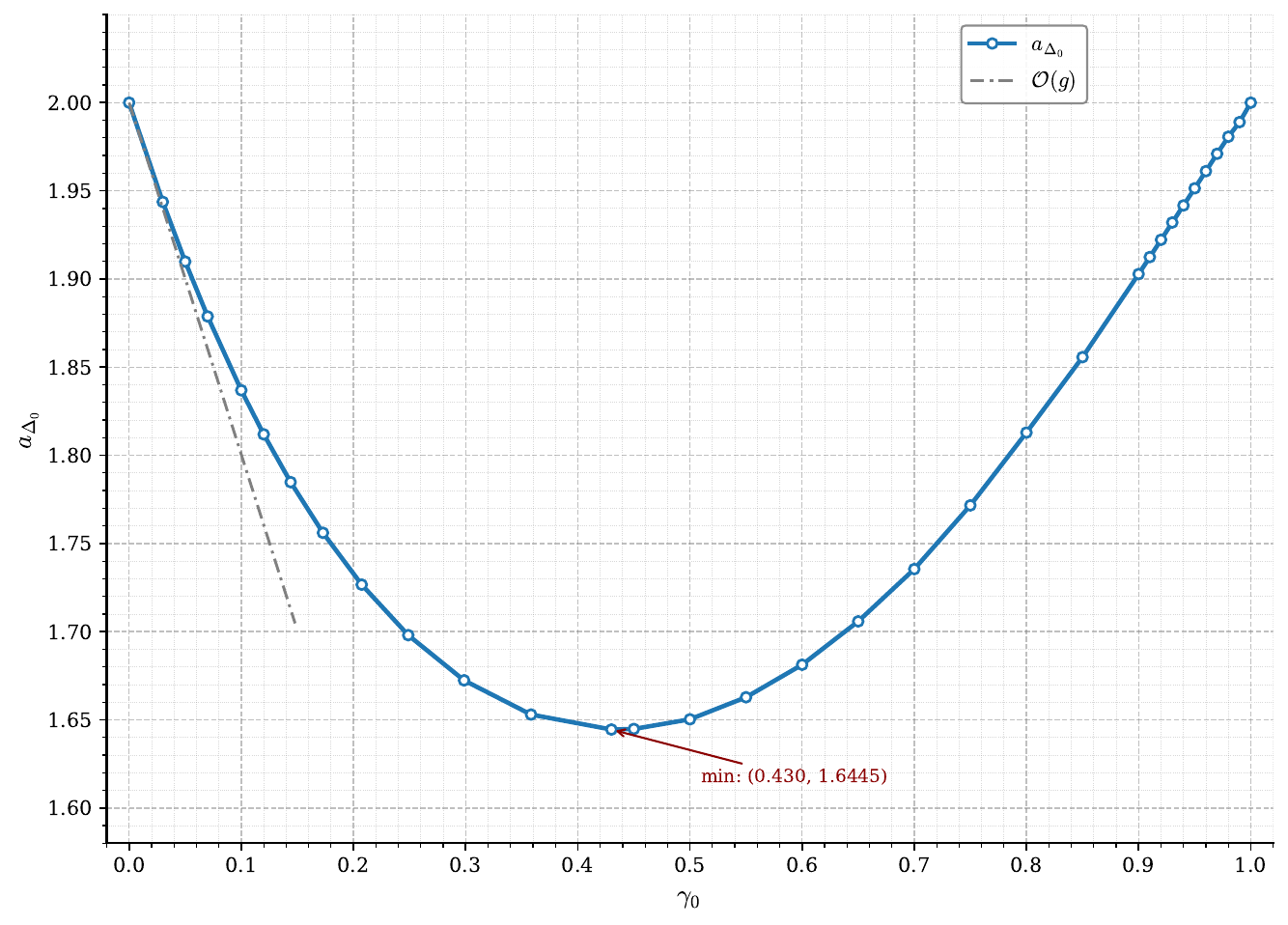}
    \caption{Variation of the first OPE coefficient $a_{\Delta_0}$ w.r.t. $\Delta_0$. The dashed line represents analytic perturbation theory computations in first order in  $g:= \Delta_1 - 2\Delta_{\phi}$. }
\label{fig:ope-coeff}
\end{figure}
In Fig. \ref{fig:ope-coeff} we show the variation of the first OPE coefficient in the minimal spectrum. As $\Delta_0\to 3$ we find $a_{\Delta_0}\to 2$ which is the respective GFF fermion value. The plot can be compared to Fig 5 of \cite{Zan:2019fkw} which shows the same OPE coefficient but from OPE maximization. Our plot has no visible difference from theirs. An interesting feature of the plot is how $a_{\Delta_0}$ reaches a minima close to $\Delta_0=2.45$ where its value is $(a_{\Delta_0})_{\text{min}}\approx 1.645$. \AK{Flow of $\Delta_1$.}
\begin{figure}
    \centering
\includegraphics[scale=0.5]{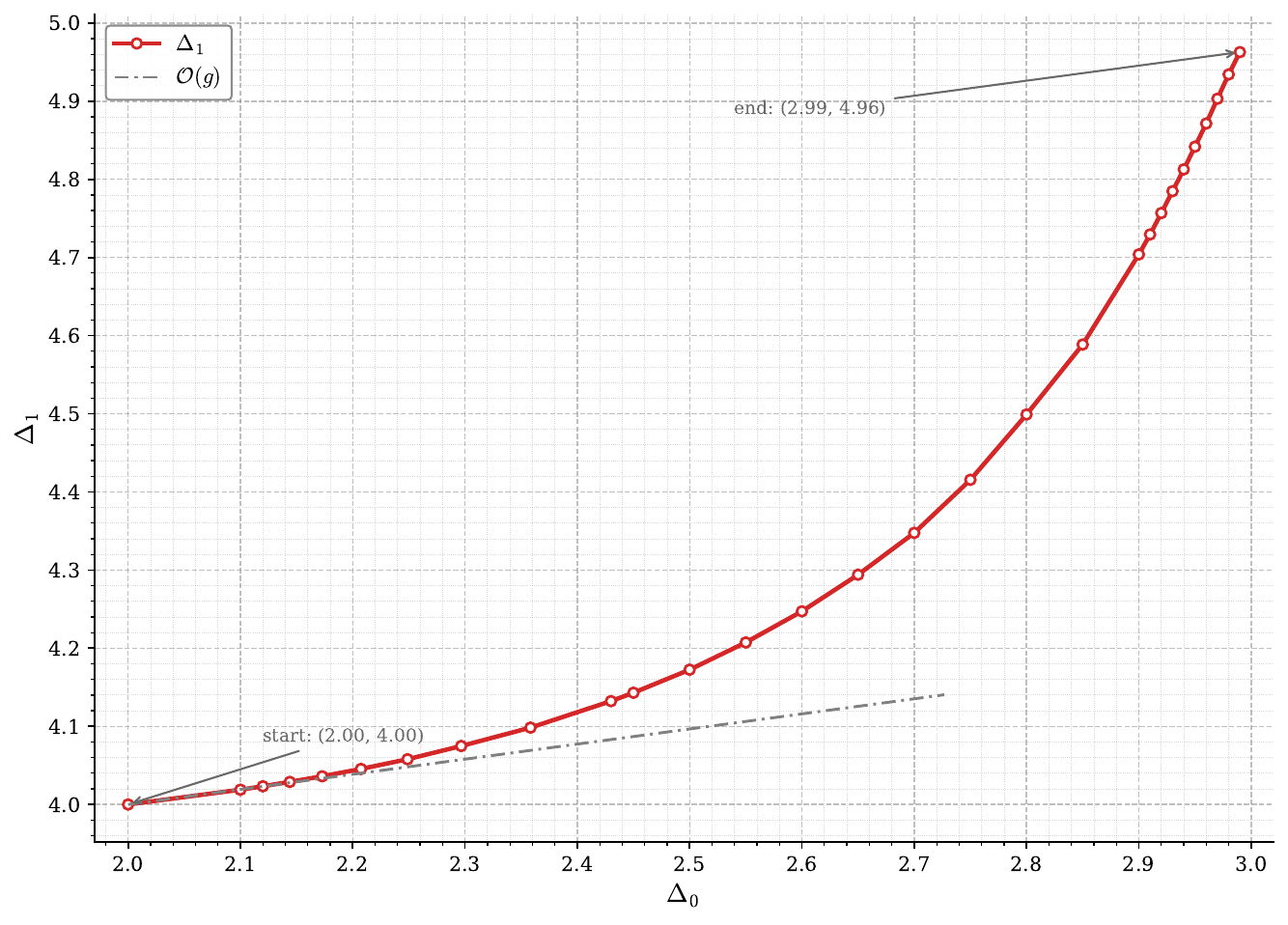}
    \caption{Variation of the second operator dimension $\Delta_1$ w.r.t. $\Delta_0$. Similar to Fig: \ref{fig:ope-coeff}, the dashed line represents analytic perturbation theory computation up to first order in  $g:= \Delta_1 - 2\Delta_{\phi}$. }
\label{fig:Delta1}
\end{figure}

In Fig. \ref{fig:Delta1} we show the specific variation of the dimension $\Delta_1=4+\gamma_1$ in the minimal spectrum flow. It reaches $\Delta_1=4.967$ for $\Delta_0=2.99$. It can be compared to the inset of Fig 5 of \cite{Zan:2019fkw} for the corresponding operator from OPE maximization extremal spectrum. Once again there is no visible difference. 

\subsection{Convergence of results}\label{error-section}

Let us show how stable our results are. The minimization problem set up in section~\ref{GD-intro} is subject to the following three modifications that can affect the results significantly:
\begin{enumerate}
\item Weights $v_n$ and $w_m$ in the loss $\mathcal{F}$.
\item Choice of initial parameters.
\item Number of sum rules $n_{\text{max}}$.
\end{enumerate}
We will choose $\gamma_0 = 0.9$ and show how the minimal spectrum changes as we vary the above. Below we will focus on only the corrections which are $\sim O(10^{-5})$ or above. This is the smallest precision that the interpolated sum rules can accurately capture. 
We find that other parameters, such as the width of each iteration in the gradient descent (`learning rate'), do not affect the results as significantly.

\begin{figure}
    \centering
\includegraphics[scale=0.4]{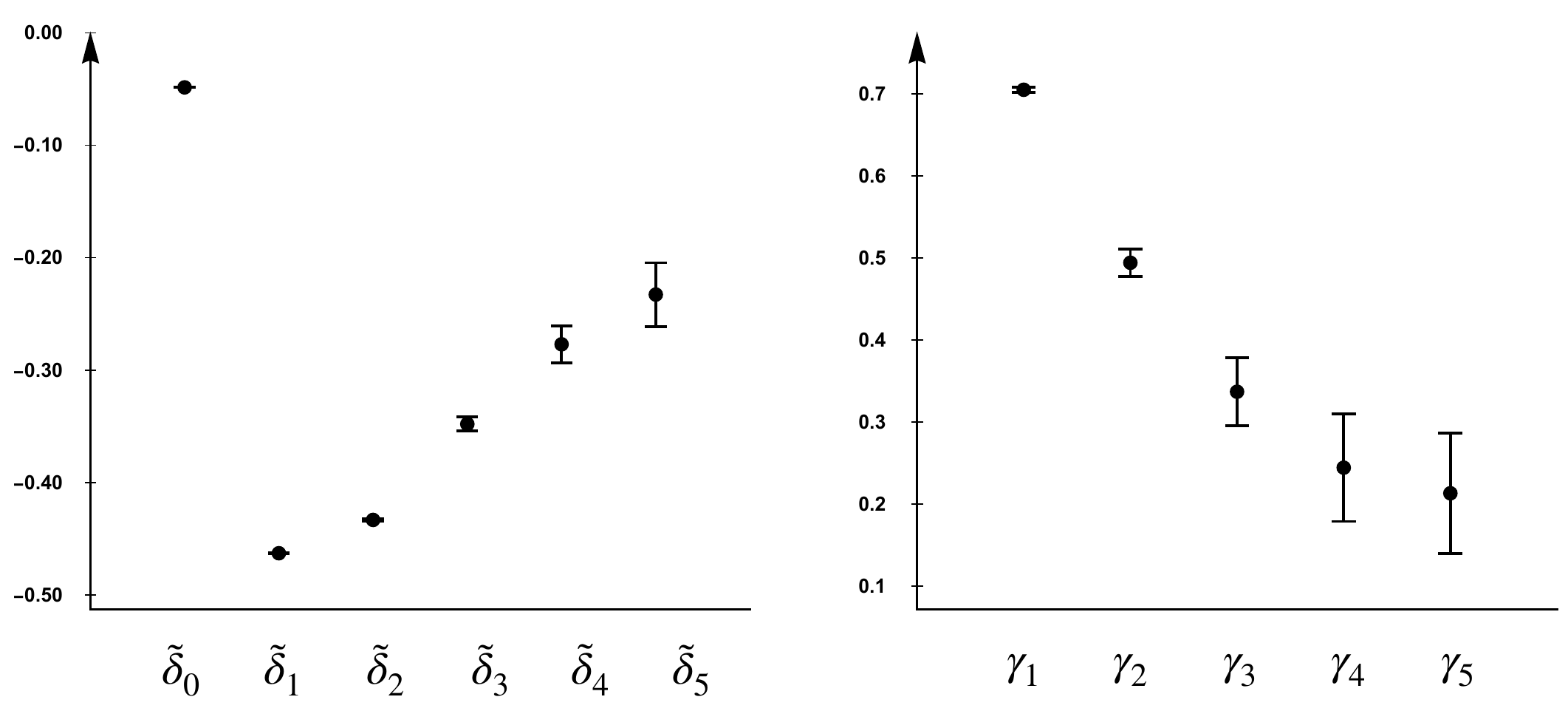}
    \caption{OPE data for various choices of $F_{\text{ref}}$ and initial values. Left: The dots represent mean of $\tilde{\delta}_n:=\delta_n/a_{n}^b$ and vertical bars represent their standard deviations. Right: The dots and vertical bars denote the mean and standard deviation for $\gamma_n$. }
\label{fig:error1}
\end{figure}

\paragraph{Weights in the loss function.} Weights refer to the quantities $v_n$ and $w_m$ in the definition \eqref{F-loss}. As we discussed in the text around \eqref{alphabeta-small-gamma}, $v_n$ and $w_m$ should increase with $n$ and $m$, since the associated sum rules get smaller when evaluated at the initial values for every gradient descent run. We proposed them to be proportional to $(\alpha_n)^{-1}$ and $(\beta_m)^{-1}$ respectively - which roughly amounts to exponential growths with $n$ and $m$. But the exact values of $v_n, w_m$ can still be chosen randomly after accounting for this exponential growth. We make a simple choice to fix them:
\begin{equation}
v_n = w_n = \left(F_{\text{ref}}\right)^{-n}.
\end{equation}
We work with $F_{\text{ref}}$ for various values from $0.25$ to $0.5$. We find this window to be the best in terms of  finding convergent results, and also being compatible with our earlier proposal.


\paragraph{Choice of initial values.} Our choice of warm starting  each gradient descent run was specified in \eqref{initial}-\eqref{initial-2}. But we want to see how sensitive our results are to that choice of warm start. So we made a number of runs with some arbitrariness in the initialization process as follows. For $\gamma_0=0.9$ we vary the initial $\gamma_n, \delta_n$ for $n=0,\cdots 4$  within the error margins of their respective final values of the $\gamma_0=0.85$ run. E.g. if we have determined $\gamma_1=0.587(4)$ for the $\gamma_0=0.85$ we start the $\gamma_0=0.9$ run with $\gamma_1=0.588, 0.583, 0.585$, etc.  The other variables, i.e. $\gamma_n, \delta_n$ for $n = 5, \ldots, 8$ come out to be too small in the $\gamma_0=0.85$ run - comparable to or smaller than their analytic values \eqref{eq:gffcorrections}. We fix them to some random initial values between 0 and 0.1. \\ \\
We can actually choose larger fluctuations  or any other convention of fluctuations for fixing initial values, but they should end up near the same minima. For larger deformations (i.e. $\gamma_0$), locating the minima is very sensitive to the initial values and large fluctuations may lead to unphysical local minima - which are easily identifiable by significantly larger values of $\mathcal{F}$. We checked that the choice of initial value fluctuations described above was consistent with finding the same minima.\\ \\ 
The mean and standard deviations for each anomalous dimension $\gamma_n$ and the GFF normalized OPE coefficient corrections $\tilde{\delta}_n:=\delta_n/a^b_n$ obtained from the runs with  various choices of weights and warm starts are summarized in Fig. \ref{fig:error1}. In all these cases we have worked with $n_{\text{max}}=8$ i.e. nine $\alpha$ and eight $\beta$ sum rules. 

\begin{table}[h]
\centering
\begin{tabular}{|c|c|c|c|c|c|}
\hline
  & $n_{\text{max}}=4$ & $n_{\text{max}}=5$ & $n_{\text{max}}=6$ & $n_{\text{max}}=7$ & $n_{\text{max}}=8$ \\
\hline
$\Delta_0$  & $2.9000$ & $2.9000$ & $2.9000$ & $2.9000$ & $2.9000$ \\
$\Delta_1$ & $4.6998$ & $4.7114$ & $4.7071$ & $4.7042$ & $4.7045$ \\
$\Delta_2$  & $6.462$ & $6.5292$ & $6.5067$ & $6.4902$ & $6.4916$ \\
$\Delta_3$  & $8.256$ & $8.4408$ & $8.3720$ & $8.3270$ & $8.3295$ \\
$\Delta_4$  & $10.086$ & $10.5799$ & $10.3213$ & $10.2274$ & $10.2263$ \\
$\Delta_5$  & $11.991$ & $12.0006$ & $12.4526$ & $12.1918$ & $12.1656$ \\
\hline
$a_{\Delta_0}$  & $1.9029$ & $1.9040$ & $1.9033$ & $1.9028$ & $1.9029$ \\
$a_{\Delta_1}$  & $0.6441$ & $0.6467$ & $0.6450$ & $0.6439$ & $0.6440$ \\
$a_{\Delta_2}$  & $0.13621$ & $0.1344$ & $0.1347$ & $0.1353$ & $0.1352$ \\
$a_{\Delta_3}$  & $0.0227$ & $0.02074$ & $0.0210$ & $0.0216$ & $0.0215$ \\
$a_{\Delta_4}$  & $0.00311$ & $0.00301$ & $0.00258$ & $0.00277$ & $0.00276$ \\
$a_{\Delta_5}$  & $-0.00077$ & $0.04241$ & $0.00028$ & $0.00030$ & $0.00030$ \\
\hline
\end{tabular}
\caption{Minimal OPE data for different choices of $n_{\text{max}}$.}\label{convergence}
\end{table}
\begin{figure}
    \centering
\includegraphics[scale=0.4]{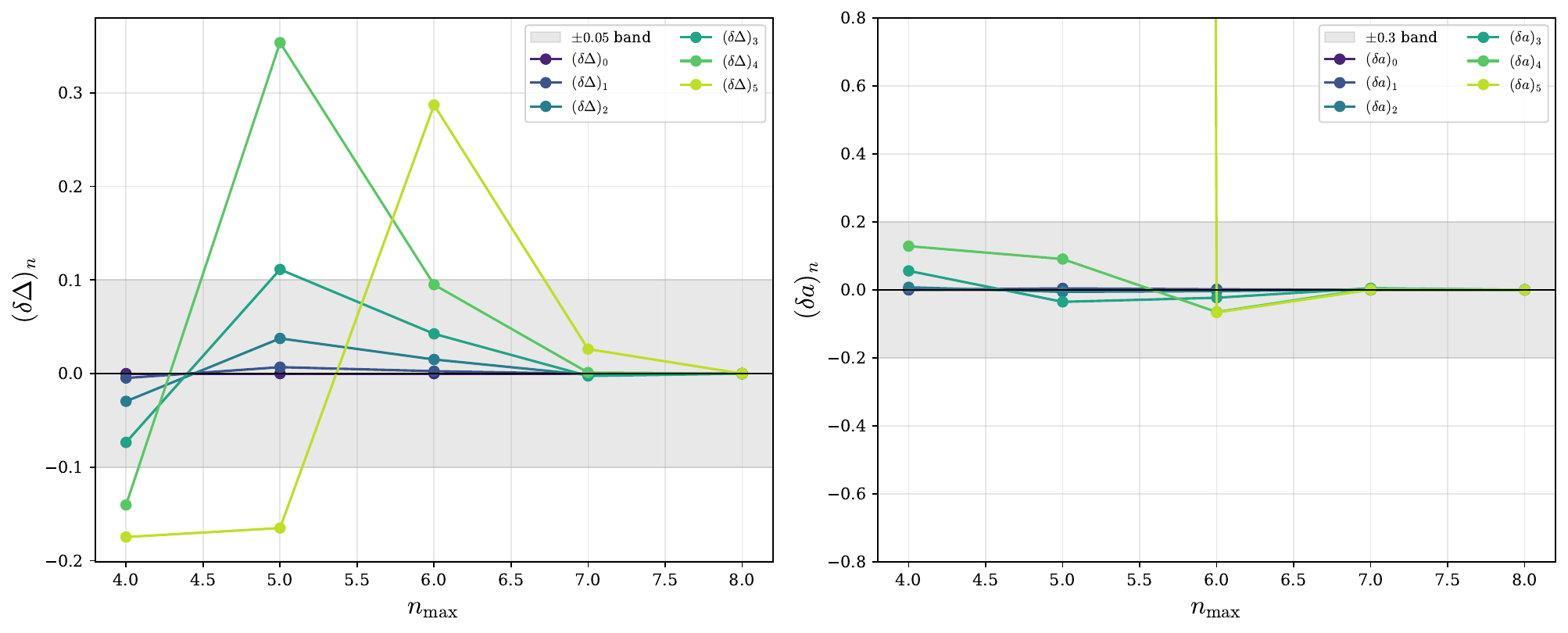}
    \caption{\textit{Left:} convergence of dimensions w.r.t. $n_{\text{max}}$. Here $(\delta\Delta)_n:= \Delta_n(n_{max})- \Delta_n(8)$. \textit{Right:} convergence of  OPE coefficients. 
    Here $(\delta a)_n:= \frac{a_{\Delta_n}(n_{max})- a_{\Delta_n}(8)}{a_{\Delta_n}(8)}$. 
    }
\label{fig:error1}
\end{figure}

\paragraph{Number of sum rules.} Although results are expected to be more accurate as we increase the number of sum rules, it is important to understand how fast the results converge as we add those constraints. So we made separate runs for $n_{\text{max}} = 4, 5, 6, 7, 8$. The summary is shown in Table \ref{convergence}. For these results we chose $F_{\text{ref}}=0.4$ and our warm start convention was according to \eqref{initial}-\eqref{initial-2}.\\ \\
We see that $\Delta_n$ and $a_{\Delta_n}$ for lower $n$'s converge faster than the higher ones. 
In Fig. \ref{fig:error1}, we try to demonstrate the convergence of the scaling dimensions $\Delta_n$ and their corresponding OPE coefficients $a_{\Delta_n}$ with respect to their mean values. For  $n_{\text{max}}=6,7,8$  most of the OPE data corresponding to $n\leq 5$ are reasonably stable.\\ \\
The final mean and standard deviations can be obtained by simultaneously varying the weights $v_n, w_m$ and the initial values in the way discussed earlier in this subsection, and for $n_{\text{max}}=6,7,8$. They  are as follows:
\begin{align}
\Delta_0 &= 2.9 \,\, \textit{(input)}\,, & a_{\Delta_0} &= 1.9031(3)\,, \label{gamma0.9-1}\\
\Delta_1 &= 4.704(3)\,, & a_{\Delta_1} &= 	0.6444(4)\,, \label{gamma0.9-2}\\
\Delta_2 &= 6.49(2)\,, & a_{\Delta_2} &= 	0.1349(3)\,, \label{gamma0.9-3}\\
\Delta_3 &= 8.33(4)\,,  & a_{\Delta_3} &= 0.0213(2)\,, \\
\Delta_4 &= 10.24(6)\,, & a_{\Delta_4} & =  0.00267(9) \,, \label{gamma0.9-4}\\
\Delta_5 &= 12.21(7)\,, & a_{\Delta_5} & =  0.00028(2) \label{gamma0.9-5}\,.
\end{align}

\subsection{Free boson deformation with an extra state}\label{newstate}

So far the only deformation to GFF has been a nonzero value of $\gamma_0$. Let us now add an extra operator in the OPE with dimension $\Delta=\Delta_\phi=1$ identical to the external dimension. We call its OPE coefficient $a_\phi$ and keep it arbitrary. 
This construction corresponds to the external field ($\phi$) appearing in its own ($\phi\times \phi$) OPE, e.g. when a $\Phi^3$ coupling is added to the action of a GFF field $\Phi$. \footnote{One would think that another example of this setup is when the external field is dual to a composite (e.g. $\Phi^2$) of a GFF field $\Phi$, but the OPE data in such cases is very specific and not the deformation of a single double twist operator tower.} \\ \\
We start with defining a loss function identical to \eqref{F-loss} but the sum over $\Delta$ running over 
\be
\Delta \in (0,1,\{\Delta_n\})\,.
\ee
We will look for solutions of the form $\Delta_n=\Delta_n^b+\gamma_n$ and $a_{\Delta_n}=a_n^b+\delta_n$. 
There are two inputs now: $a_\phi$ and $\gamma_0$. We choose to fix $\gamma_0=0.1$ and will gradually vary $a_\phi$ from $0$ to $1$ and then from $0$ to $-1$. The $(\gamma_0=0.1, a_\phi=0)$ solution was already determined to high accuracy from the numerics of section \ref{GD-no-new}. It is simple to deform $a_\phi$ from there in either direction in small steps. 

\begin{figure}
    \centering
    \includegraphics[scale=0.6]{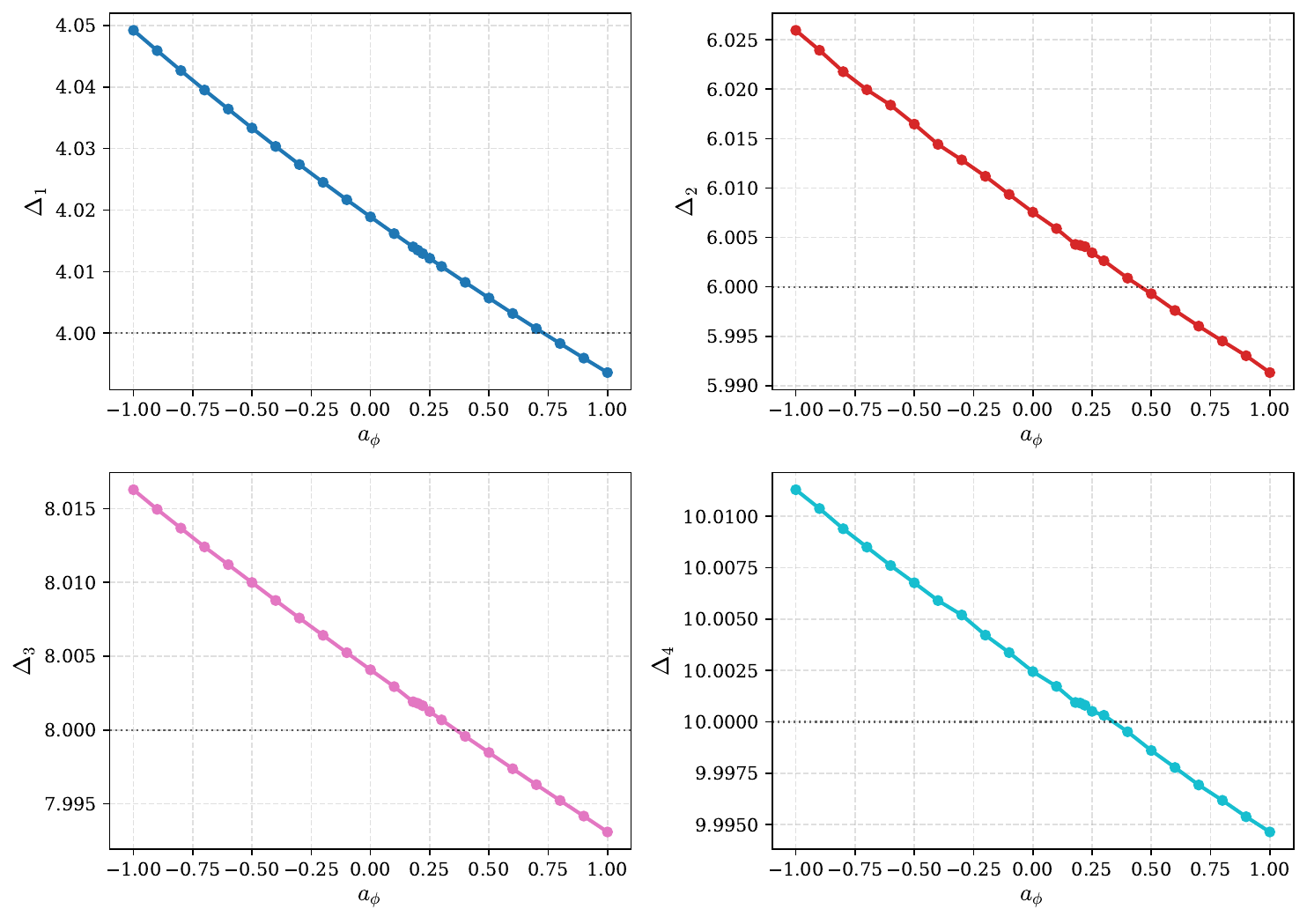}
    \caption{Variation of scaling dimensions $\Delta_n$ of GFF operators $\Delta=4,6,8,10(\Delta_{\phi}=1)$ respectively as the OPE coefficient $(a_{\phi})$ of new operator $\Delta=1$ is varied from $-1$ to $1$ keeping $\gamma_0 = 0.1$.}
    \label{fig:aphigamma}
\end{figure}

\begin{figure}
    \centering
\includegraphics[scale=0.6]{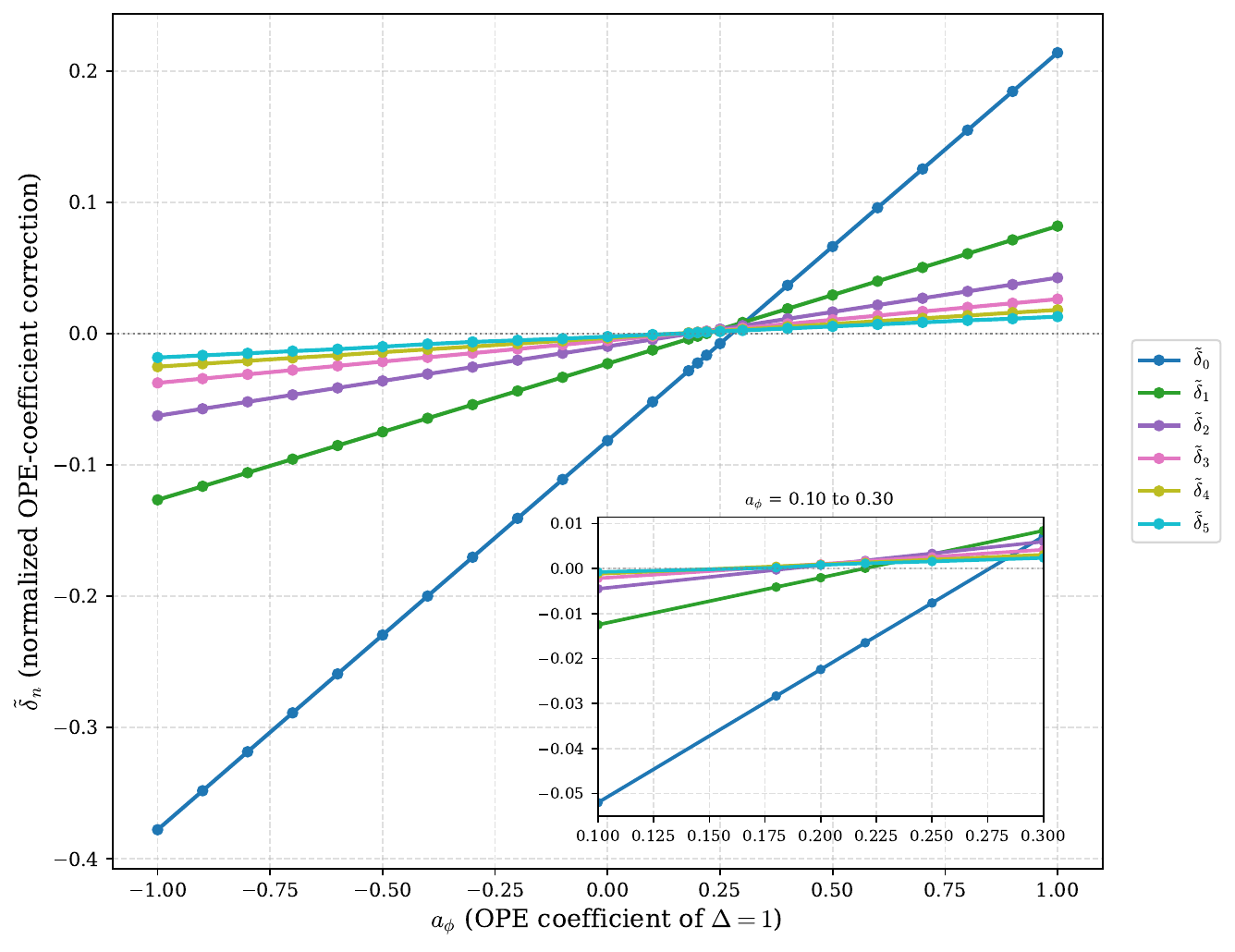}
    \caption{The variation of normalized OPE corrections $\tilde{\delta}_n$ w.r.t. OPE coefficient $(a_{\phi})$ of new operator $\Delta = 1$. }
\label{fig:aphidelta}
\end{figure}

For this example we worked with $n_{\text{max}}=6$. Adding higher sum rules does not change the minimal spectrum but only slows down the minimization. In Fig. \ref{fig:aphigamma} we show the variation of the dimensions $\Delta_n$, $n=1,2,3,4$ with $a_\phi$. In Fig. \ref{fig:aphidelta} we show the variation of the normalized OPE coefficient corrections $\tilde{\delta}_n:=\delta_n/a^b_n$. We do not show the results for  $\gamma_{n\ge 5}$ and $\tilde{\delta}_{n\ge 6}$ as they are too small. 

Let us point out some noteworthy observations. The solutions corresponding to negative $a_\phi$ are by construction non-unitary. However for each solution the OPE data for $\Delta_n$ gradually approach the free spectrum as $n$ increases and hence follow the pattern of asymptotic freedom of \cite{Ghosh:2025sic}. Following their paper it is tempting to suggest that each such spectrum would correspond to an OPE maximization problem - for the $\Delta_0$ operator. But for a non-unitary theory this would be hard to prove. Indeed we are free to insert another operator with dimension $\Delta_0'\approx \Delta_0$ with a negative OPE coefficient such that $a_{\Delta_0}$ is allowed to be higher than what we find. However such a solution has a different counting than GFF (see the discussion below \eqref{gff defo}) as it has an extra operator over the double twist tower. To see such a solution in the truncated PB  we may fix that extra operator in the spectrum along with the one at $\Delta=\Delta_\phi$ with varying OPE coefficients - to connect smoothly to GFF. This shows how different deformations span the space of solutions - including nonunitary ones. 





\section{Truncated bootstrap with mixed correlators}\label{mixed-PB-GD}
So far we only considered the constraints from a single correlator, and obtained solutions involving 
corrections to only the double twist operator tower in the $\phi\times \phi$ OPE. 
Now we will add constraints from more than one correlator in the truncated PB. This will introduce operator(s) of a higher twist family in the same OPE.

\paragraph{OPE and correlators:}
Let us take a deformation of the bosonic GFF that preserves a $Z_2$ symmetry. We may denote $\Phi_1:=\phi$ as the lightest $Z_2$ odd operator with $\Delta_\phi=1$ - the deformed GFF boson. We also denote as $\Phi_2$  the lightest $Z_2$ even operator that appears in $\Phi_1\times\Phi_1$ OPE.
Now let us focus on the correlators with only $\Phi_1$ and $\Phi_2$ as external fields - there are three of them:
\begin{align}
\mathcal{G}^{1111}(\{x_i\})=&\langle\Phi_1(x_1)\Phi_1(x_2)\Phi_1(x_3)\Phi_1(x_4)\rangle\,,\label{G-1}\\
\mathcal{G}^{1122}(\{x_i\})=&\langle\Phi_1(x_1)\Phi_1(x_2)\Phi_2(x_3)\Phi_2(x_4)\rangle\,,\label{G-2}\\
\mathcal{G}^{2222}(\{x_i\})=&\langle\Phi_2(x_1)\Phi_2(x_2)\Phi_2(x_3)\Phi_2(x_4)\rangle\,.\label{G-3}
\end{align}
For $\mathcal{G}^{1111}$ the relevant OPE is given by:
\be\label{11ope}
\Phi_1\times \Phi_1\sim \textbf{1}\,, \mathcal{O}_n\,, \, \mathcal{O}'_m\,.
\ee
Here $\mathcal{O}_n$ and $\mathcal{O}'_m$ are  $Z_2$ even operators, with the respective dimensions $\Delta_n=2\Delta_\phi+2n+\gamma_n$ and $\Delta_m'=4\Delta_\phi+2m+\gamma_m'$. Note that $\Phi_2=\mathcal{O}_0$. We consider a truncated spectrum, i.e.  \ $n=0,1,\cdots , n_{\text{max}}\,$ and $m=0,\cdots , m_{\text{max}}$\,. The $n$ and $m$ respectively label the (deformed) double twist and quartic twist families. One could include other $Z_2$ even multi-twist operators too but the mixed correlator system we consider (i.e. \eqref{G-1}-\eqref{G-3}) will not detect them. \\ \\
For the other two correlators the following two OPEs are involved:
\begin{align}
    &\Phi_2\times \Phi_1 \, \sim \,  \phi\,, \, \mathcal{O}_k''\,,\label{12ope}\\
    &\Phi_2\times \Phi_2 \, \sim \textbf{1}\,, \mathcal{O}_n\,, \, \mathcal{O}'_m\,. \label{22ope}
\end{align}
Here $\mathcal{O}_k''$ are $Z_2$ odd operators with the dimensions $\Delta_k''=3\Delta_\phi+k+\gamma_k''$. They belong to the triple twist operator family. Notice that the operator $\phi$ itself appears in the $\Phi_2\times \Phi_1 $ OPE  - this is easy to see at the exact GFF point where $\Phi_2=\mathcal{O}_0=\phi^2$. Also the spectra of $\Phi_1\times\Phi_1$ and $\Phi_2\times\Phi_2$ are the same since both are $Z_2$ even OPEs. Recall that for mixed correlators every operator has two labels, $\Delta$ and $P=\pm 1$ (parity). In the above OPEs $\mathcal{O}_n$ and $\mathcal{O}'_m$ have $P=1$ whereas $\mathcal{O}''_k$ have $P=(-1)^k$. Once again the multi-twist families we need to consider are limited by the system of correlators we choose.  

We will denote the OPE coefficient  of  $\mathcal{O}$ with labels $(\Delta,P)$ in $\mathcal{G}^{ijkl}$ in the channel $c=(ij,kl)$ as $a^{ij,kl}_{\Delta,P}$ or simply $a^{ij,kl}_{\mathcal{O}}$. 
This means one has $a^{ij,kl}_{\Delta,P}=\lambda^{ij}_{\Delta,P}\lambda^{kl}_{\Delta,P}$. 
Since an operator can appear in more than one OPE its 4-point OPE coefficients will be related by $(a^{ij,kl}_{\Delta,P})^2=a^{ij,ik}_{\Delta,P}a^{kl,kl}_{\Delta,P}$. More specifically we have:
\begin{align}\label{ope-rel}
(a^{11,22}_{\Delta,P})^2=a^{11,11}_{\Delta,P}\,a^{22,22}_{\Delta,P}\,.
\end{align}
Furthermore, since $\lambda^{11}_{\Phi_2}=\lambda^{12}_{\Phi_1}$ one has 
\be\label{ope-rel2}
a^{11,11}_{\Phi_2}=a^{12,21}_{\Phi_1}\,.
\ee

\paragraph{Setting up loss functions:}
We will use PB sum rules for the multiple correlator system. Since $\mathcal{G}^{1111}$ and $\mathcal{G}^{2222}$ have identical external operators the sum rules are simply the ones of sec. \ref{ordinaryPB}. Let us introduce the following loss functions
\begin{align}
    \mathcal{F}^{1111}=\sum_{n=0}^{n_\text{max}} v_n\Big[ \sum_{\Delta}a^{11,11}_{\Delta}\alpha^{1111}_{n}(\Delta)\Big]^2\, + \, \sum_{n=1}^{n_\text{max}} w_m\Big[ \sum_{\Delta}a^{11,11}_{\Delta}\beta^{1111}_m(\Delta)\Big]^2\,,\\
    \mathcal{F}^{2222}=\sum_{m=0}^{m_\text{max}} v'_n\Big[ \sum_{\Delta}a^{22,22}_{\Delta}\alpha^{2222}_{n}(\Delta)\Big]^2\, + \, \sum_{m=1}^{m_\text{max}} w'_m\Big[ \sum_{\Delta}a^{22,22}_{\Delta}\beta^{2222}_m(\Delta)\Big]^2\,.
\end{align}
The sum over $\Delta$ in the above runs over all $Z_2$ even operators which are also parity even ($P=+1$) (so we ignored parity dependence in the sum rules).
The sum rules $\alpha^{iiii}$ and $\beta^{iiii}$ are simply the $\alpha, \beta$ sum rules from eq. \eqref{sumrules} but evaluated for the external dimensions $\Delta_\phi=1$ and $\Delta_\phi=2+\gamma_0$ for $i=1$ and $i=2$ respectively. The upper limits $n_{\text{max}}$ and $m_{\text{max}}$ should be the same as the truncation number of $\mathcal{O}_n$ and $\mathcal{O}'_m$ families respectively - this way each sum rule corresponds to an OPE data if we are close to GFF.

For the mixed correlator $\mathcal{G}^{1122}$ there are four sets of sum rules: $\alpha^{1122}_{11,n}$, $\alpha^{1122}_{22,m}$, $\alpha^{1221}_k$ and $\beta^{1221}_k$ as we described in sec. \ref{mixed-PB} (see eq. \eqref{eq:mixedsumrules1}, \eqref{eq:mixedsumrules2}). Recall that the superscripts indicate which channel the mixed Polyakov block was expanded in (e.g. $\alpha^{1122}$ was obtained by expanding in the $(11,22)$ channel), see eq. \eqref{eq:1423channel}, \eqref{eq:1234channel}. Accordingly we introduce another loss function:
\begin{align}
&\mathcal{F}^{1122}=\sum_{p=0}^{p_\text{max}} w^{(1)}_p\Big[ \sum_{\Delta,P}\sum_{c}a^{c}_{\Delta,P}\,\alpha^{1122,c}_{11,p}(\Delta,P)\Big]^2\, + \, \sum_{q=0}^{q_\text{max}} w^{(2)}_q\Big[ \sum_{\Delta,P}\sum_{c}a^{c}_{\Delta,P}\,\alpha^{1122,c}_{22,q}(\Delta,P)\Big]^2\,\nonumber\\
&+\, \sum_{k=0}^{k_\text{max}} w^{(3)}_k\Big[ \sum_{\Delta,P}\sum_{c}a^{c}_{\Delta,P}\,\alpha^{1221,c}_{k}(\Delta,P)\Big]^2\,+\, \sum_{k=2}^{k_\text{max}} w^{(4)}_k\Big[ \sum_{\Delta,P}\sum_{c}a^{c}_{\Delta,P}\,\beta^{1221,c}_k((\Delta,P)\Big]^2\,.
\end{align}
Here $c$ runs over the channels i.e. $(11,22)$, $(12,21)$ and $(12,12)$. Recall that the OPE coefficient of each operator changes according to the channel. E.g. if a certain operator $\mathcal{O}_{\Delta,P}$ is present in the $(11,22)$ channel but not in $(12,21)$ or $(12,12)$ we should have $a_{\Delta,P}^{11,22}\neq 0, \  a_{\Delta,P}^{12,21}=a_{\Delta,P}^{12,12}=0$.  The expressions $\alpha^{1122,c}_{11,p}$, $\alpha^{1122,c}_{22,q}$, $\beta^{1122,c}_k$ and $\alpha^{1221,c}_k$ entering the sum rules were defined in eqns. \eqref{eq:1234channel}, \eqref{eq:1423channel}. Recall also the property \eqref{mixed-sumrule-property} on how each sum rule captures a certain perturbative correction. So, the number $k_{\text{max}}$ corresponds to how many $\mathcal{O}''_k$ get corrected. The numbers $p_{\text{max}}$ and $q_{\text{max}}$ are chosen according to how many OPE coefficients $a_{\mathcal{O}_n}^{11,22}$ and $a_{\mathcal{O}_m'}^{11,22}$ respectively should be corrected. Note that these three truncations can be set independent of one another.\\ \\
To bootstrap the complete mixed correlator system one has to work with the total loss function as follows
\be
\mathcal{F}^{\text{tot}}=\mathcal{F}^{1111}+\mathcal{F}^{2222}+\mathcal{F}^{1122}\,.
\ee
Consider the first two loss functions $\mathcal{F}^{1111}$ and $\mathcal{F}^{2222}$.  Although the $\Delta$'s appearing in these two are the same, the OPE coefficients entering them ($a^{11,11}_{\Delta,P}$ and $a^{22,22}_{\Delta,P}$ respectively) are different. If these OPE coefficients were independent the sum rules involved would be an underdetermined system of equations and without a unique minimal spectra. But eqn. \eqref{ope-rel} relates the two OPE coefficients to $a^{11,22}_{\Delta,P}$ which enters the third loss $\mathcal{F}^{1122}$. So only by minimizing the combined loss $\mathcal{F}^{\text{tot}}$ does one get a unique minimal spectrum.

Let us now explain its consequence.
Consider the operator $\mathcal{O}'_{0}=\phi^4$ in GFF which appears in $\Phi_2\times\Phi_2$, but not in $\Phi_1\times\Phi_1$ (i.e. $a^{11,11}_{\mathcal{O}'_{0}}=0$ in GFF, while $a^{22,22}_{\mathcal{O}'_{0}}\neq 0$). This also implies $a^{11,22}_{\mathcal{O}'_{0}}=0$ by \eqref{ope-rel}, which is consistent with the mixed correlator sum rules in $\mathcal{F}^{1122}$ with GFF data.
Now let us introduce a small deformation $\gamma_0=\Delta_0-2\Delta_\phi\neq 0$. The sum rule $\alpha^{1122}_{22,0}$ would now force us to turn on $a^{11,22}_{\mathcal{O}'_{0}}$. This would also imply $a^{11,11}_{\mathcal{O}'_{0}}\neq 0$, and hence the introduction of a new state in the $\Phi_1\times \Phi_1$ OPE. 

\paragraph{Variables and input: } We keep the simplest version of the above setup. Namely, we truncate the $\mathcal{O}_n$ and $\mathcal{O}'_m$ families at $n=4$ and $m=3$ respectively (see eqs. \eqref{11ope} -  \eqref{22ope}).  In accordance we set 
\begin{align}
&n_{\text{max}}=4\,,\nonumber\\
&m_{\text{max}}=3\,.
\end{align}
in $\mathcal{F}^{1111}$ and $\mathcal{F}^{2222}$ (we restrict the deformations, see \eqref{mixed-input} below, so that these numbers are enough for convergence). Also we truncate the $\mathcal{O}''_k$ family at $k=1$ and in $\mathcal{F}^{1122}$ set
\begin{align}
    &k_{\text{max}}=1\nonumber\\
    &p_{\text{max}}=1\nonumber\\
    &q_{\text{max}}=0\,.
\end{align}
We choose the variables to be the deformations around their GFF values. So the full set of variables is the following:
\begin{align}
    &\gamma_n, \gamma'_m, \gamma''_k, \\
    &(\delta a)^{11,11}_{\mathcal{O}}=a^{11,11}_{\mathcal{O}}-{(a^{11,11}_{\mathcal{O}})}_{\text{GFF}}\,, \ \text{ where } \mathcal{O}\in \{\mathcal{O}_n,\mathcal{O}'_m\}, \nonumber\\
    &(\delta a)^{22,22}_{\mathcal{O}}=a^{22,22}_{\mathcal{O}}-{(a^{22,22}_{\mathcal{O}})}_{\text{GFF}}\,, \ \text{ where } \mathcal{O}\in \{\mathcal{O}_n,\mathcal{O}'_m\}, \nonumber\\
    &(\delta a)^{11,22}_{\mathcal{O}}=a^{11,22}_{\mathcal{O}}-{(a^{11,22}_{\mathcal{O}})}_{\text{GFF}}\,, \ \text{ where } \mathcal{O}\in \{\mathcal{O}_n,\mathcal{O}'_m\}, \nonumber\\
    &(\delta a)^{12,21}_{\mathcal{O}}=a^{12,21}_{\mathcal{O}}-{(a^{12,21}_{\mathcal{O}})}_{\text{GFF}}\,, \ \text{ where } \mathcal{O}\in \{\phi\,, \mathcal{O}''_k\}\,.
\end{align}
Some of the variables are not independent, due to the relations \eqref{ope-rel} and \eqref{ope-rel2}. Analogous to fixing $\gamma_0$ as an input for the single correlator case, now we have a choice to fix four of the variables. We choose to fix the following variables as inputs:
\be\label{4input}
\gamma_0\,, \ \gamma''_0\,, \ (\delta a)^{22,22}_{\mathcal{O}'_0}\,, \ (\delta a)^{12,21}_{\mathcal{O}''_1}\,.
\ee
They are four in number because the mixed correlator PB sum rules are meant to bootstrap correlators of independent GFF fields. For a set of correlators like \eqref{G-1}-\eqref{G-3}, if the external operators $\Phi_1$ and $\Phi_2$ were two independent GFF fields they would allow deformations by four independent contact diagrams - one each for $\mathcal{G}^{1111}$ and $\mathcal{G}^{2222}$, and two for $\mathcal{G}^{1122}$ (see the discussion below \eqref{eq:3.1.18} and below \eqref{mixedPB1122}).

\begin{figure}
    \centering
\includegraphics[scale=0.35]{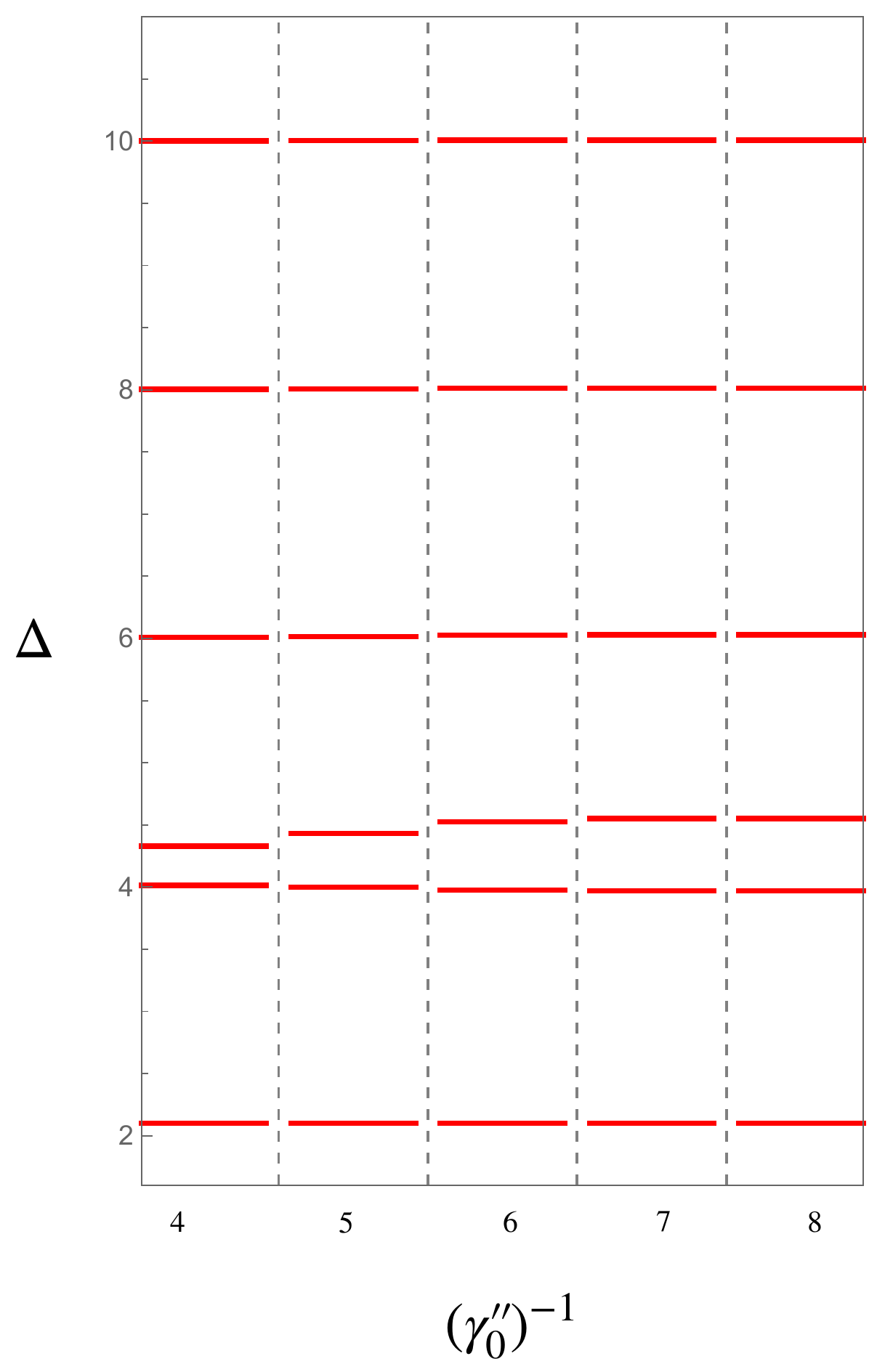}
    \caption{The spectrum of $\phi\times\phi$ OPE from mixed correlator sum rules  for different values of $\gamma_0''$. Here we chose $\gamma_0=0.1$ and $(\delta a)^{22,22}_{\mathcal{O}'_0}=0.2$.}
\label{fig:spectrum2}
\end{figure}

However we are aiming to bootstrap the deformation of the theory of a single GFF field $\Phi_1=\phi$ (and $\Phi_2=\mathcal{O}_0$ corresponds to $\phi^2$). This theory has only three independent contact diagrams formed with the vertices ${\Phi}^4, {\Phi}^6, {\Phi}^8$, where ${\Phi}$ is a massive free field in AdS$_2$. The first three parameters in \eqref{4input} fix the strengths of these contacts. To fix the last one, we observe that there is no primary operator $\mathcal{O}''_1$ with dimension $3\Delta_\phi+1$ at GFF (so $(a^{12,21}_{\mathcal{O}''_1})_{\text{GFF}}=0$). We enforce that such a state will not emerge even after deformation, so we fix
$(\delta a)^{12,21}_{\mathcal{O}''_1}=a^{12,21}_{\mathcal{O}''_1}=0$. If we consider deformations of GFF by interactions like ${\Phi}^4, {\Phi}^6$ or ${\Phi}^8$ this choice is natural. \\ \\
We fix $\gamma_0$ and $(\delta a)^{22,22}_{\mathcal{O}'_0}$ and keep $\gamma_0''$ flexible for the analysis:
\begin{align}\label{mixed-input}
&\boxed{\gamma_0=0.1\,,  \ (\delta a)^{22,22}_{\mathcal{O}'_0}=0.2\,,}\nonumber\\
&\hspace{1cm}\boxed{\mbox{$\frac{1}{8}$}\le  \gamma_0''\le \mbox{$\frac{1}{4}$}\,.}
\end{align}
We have also run the optimization with other values of $(\delta a)^{22,22}_{\mathcal{O}'_0}$. The final results were not very sensitive to those changes, so we do not report them.
The values of the $\gamma_0''$ and $(\delta a)^{22,22}_{\mathcal{O}'_0}$ are chosen in a range for which a satisfactory global minima of $\mathcal{F}^{\text{tot}}$ can be obtained. This means  in that range every sum rule in $\mathcal{F}^{\text{tot}}$ is reduced by a large factor compared to its value for the initial parameters. 
\begin{figure}
    \centering
\includegraphics[scale=0.25]{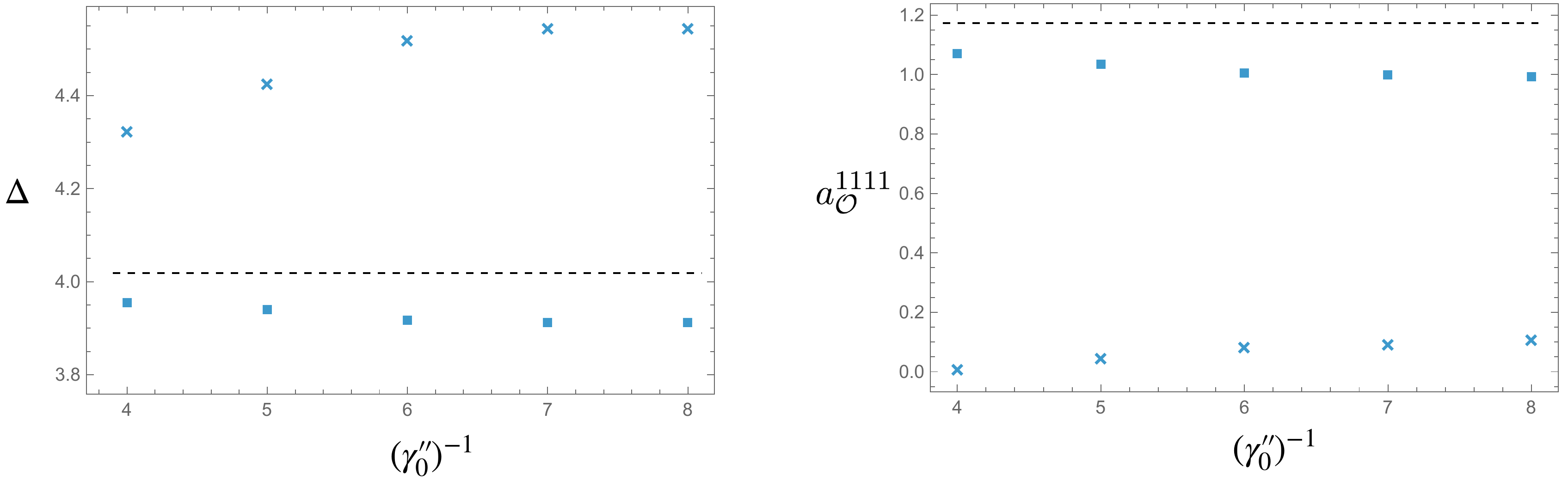}
    \caption{The dimension (left) and OPE coefficients (right) of the operators $\mathcal{O}_0$ (square dots) and $\mathcal{O}'_1$ (crosses) in $\phi\times\phi$ OPE for different $\gamma_0''$ values, with $\gamma_0=0.1$ and $(\delta a)^{22,22}_{\mathcal{O}'_0}=0.2$. The dashed lines in both the diagrams represent the respective OPE data for $\mathcal{O}_0$ from the single correlator sum rules.}
\label{fig:zoomed}
\end{figure}
\paragraph{Results: } We used two different data sets for initialization: one was simply the free spectrum, i.e. all the variables set to zero. The second was the minimal spectrum of $\gamma_0=0.1$ from the single correlator case, i.e. from Table \ref{gam0.1}. Both warm starting points converge to the same data that minimize $\mathcal{F}_{\text{tot}}$. 

In Fig. \ref{fig:spectrum2} we show the spectrum of dimensions in $\Phi_1\times \Phi_1$ OPE. We show the results for five different values of $\gamma_0''$ while $(\delta a)^{22,22}_{\mathcal{O}'_0}$ was held fixed at the value $0.2$.  There are two operators in the OPE with dimension close to $4$. The lower one and upper one correspond to $\mathcal{O}_1$ and $\mathcal{O}'_0$ respectively. We isolate their OPE data in Fig. \ref{fig:zoomed}. \\ \\
Recall that we limited the sum rules in $\mathcal{F}^{1122}$ with $q_{\text{max}}=0$ i.e. with a single $\alpha^{1122,c}_{22,q}$ sum rule. Of course if we increase $q_{\text{max}}$ we will see higher quartic twist operators (i.e. higher $\mathcal{O}'_m$-s) appearing in $\Phi_1\times \Phi_1$ OPE. Indeed that would  be closer to the full crossing solution of the mixed correlator setup. However we saw for the single correlator case in section \ref{GD-no-new} that the corrections to the higher OPE data (i.e. higher $n$'s) are smaller. Indeed the lower end of the spectrum does not change much as we change $n_{\text{max}}$. We expect the mixed correlator case to have a similar property for both double and quartic trace operator towers. 

However if the lower spectrum corrections come out to be large we should be careful and repeat the analysis with more sum rules. A large correction implies the minimal spectrum is not close to the initialization data (GFF or the single correlator solution) - which is our main assumption. Even holographically, if contact interactions are added to the free massive AdS$_2$ scalar theory, the quartic trace operators in $\phi\times\phi$ are expected to be suppressed in tree level and appear only at one loop \cite{Antunes:2021abs}.  For the input we chose, i.e. \eqref{mixed-input}, these expectations are not violated. But a general deformation of GFF needs a more extended mixed correlator analysis than the one presented here - which we leave for a future work. 

\section{Discussion and outlook}\label{conclusion}

We presented a formulation of the truncated bootstrap based on Polyakov bootstrap (PB) sum rules. The main difference of the truncated PB with other truncation schemes is that it relies only on how a crossing solution is related to the generalized free field (GFF) theory by some deformation.
General non-perturbative solutions are obtained by introducing a deformation perturbatively, increasing it in small steps and requiring that all sum rules be solved simultaneously at each step.
The step-wise search is what gives us the control in scanning over a huge multidimensional space of OPE data to find bootstrap solutions. We numerically implement this by minimizing a sum rule loss function via gradient descent.\\ \\
Our numerical method does not rely on unitarity-based positivity of OPE coefficients. 
However, due to the nature of our truncation scheme and  the asymptotic freedom conjecture of \cite{Ghosh:2025sic}, many of the solutions in the single correlator setting can be identified as extremal. But extremality does not set the limit of the truncated PB. This is clear from our findings with nonunitary deformations. 
Our mixed correlator explorations enrich the crossing solutions by letting different multi-twist operator towers exist simultaneously. 
This points at how arbitrary the form of the spectrum can be as the deformations need not be restricted to light operators only. 
\\ \\
Let us address, point-wise, the future prospects of this work:

\begin{enumerate}
\item \textbf{Higher point functions:} General $n$-point correlation functions form an exciting and active corner of the bootstrap \cite{Poland:2023vpn, Harris:2024nmr, Pal:2023kgu}, as they capture the OPE data of infinite 4-point functions. However the simplest higher point, $n=5$, correlator lacks positivity of OPE coefficients in its block decomposition.  The truncated PB approach is, therefore, a natural playground for 5-point functions. In 1d, PB sum rules for 5-point correlators were obtained in \cite{Antunes:2025vvl}.

\item \textbf{Boundary and defect CFTs:} It is well known that the crossing symmetry formulation of boundary or defect correlation functions lacks positivity, which makes it hard and unreliable to set up positive semidefinite bootstrap \cite{Liendo:2012hy}. However, often one has good analytic control over such theories from perturbation theory. In \cite{bcft-paper} the present formulation will be extended to BCFTs. It will be interesting to explore more general BCFT examples and defects in the future.

\item \textbf{Higher dimension general CFTs:} In higher dimensions, the analogues of PB sum rules are the CFT dispersive sum rules \cite{caron2021dispersive, Carmi:2020ekr, Bianchi:2021piu} that have been related to CFT dispersion relations \cite{Penedones:2019tng, Carmi:2024tmp}. The crossing symmetric dispersion relation (CSDR) makes the connection to PB more concrete \cite{gopakumar2021crossing}. Furthermore, in integer dimension a powerful alternate formulation of analytic functionals exists -- there are the ``product functionals'' originally proposed in \cite{Paulos:2019gtx} and whose applications were extensively shown in \cite{Ghosh:2023onl, Ghosh:2026dse}.

\item \textbf{AdS QFTs and S-Matrix bootstrap:} Placing a QFT in AdS background is a useful way to understand its S-matrices and RG flow \cite{Antunes:2021abs,Antunes:2024hrt,Antunes:2025iaw}, and a natural application for the PB approach. An interesting but simple extension of the truncated PB is to incorporate conditions that can isolate such  holographic theories, such as bulk locality \cite{Levine:2023ywq, Levine:2024wqn} or integrability \cite{Cavaglia:2022yvv}.

Perhaps the most exciting future prospect is an alternate formulation for bootstrapping flat space S-matrices. An amplitude satisfying all bootstrap axioms, e.g.\ the Veneziano-Shapiro amplitude or its generalizations \cite{Bocchia:2026kew, Shao:2026akl}, could be used as the seed theory for deformations in a truncated PB style approach. Although the exact analogues of PB sum rules are not known for flat space scattering amplitudes, it should be possible to reformulate our working philosophy with CSDR for scattering and its variants \cite{Sinha:2020win, Bhat:2025zex}.

\item \textbf{NN based reconstruction of CFT data:} Neural network (NN) based algorithms have recently emerged as an effective tool for bootstrapping CFT correlators \cite{Ghosh:2026jbw}. Simple feed-forward NNs trained on crossing symmetry, some minimal input on spectral data and the principle of smoothness of conformal correlators lead to remarkable reconstruction of correlation functions \cite{Ghosh:2026xnp,Ghosh:2026efw}. Dispersive sum rules, on the other hand, are consistency conditions on  correlator with given asymptotics and analytic structure. The predictive power of NN based reconstruction could be expected to improve by using dispersive constraints to regulate the space of functions the network is allowed to learn. This could lead to more accurate predictions and a better analytic handle on CFT data and observables.

\end{enumerate}

\paragraph{Acknowledgement: } We thank Ant\'onio Antunes, Diptarka Das, Kausik Ghosh, Koushik Ray, Aninda Sinha, Philine van Vliet for discussions. We thank Ant\'onio Antunes and Philine van Vliet  for comments on the draft. We acknowledge Claude (Anthropic) and ChatGPT (OpenAI) for assistance with numerical coding. AK is supported by ANRF grant ANRF/ARG/2025/001338/PS. SL’s computational work was carried out on equipment funded by Beijing Natural Science Foundation grant no. IS23010\,.

\pagebreak

\appendix
\section*{Appendix:}
\section{Witten Diagrams and Mellin Space}
\label{app:Witten}
The sum rules we used extensively in the main text are derived from 1d CFT Polyakov blocks. In this appendix we give the details of the Witten diagrams which are the main components of the Polyakov blocks. \\ \\
Let us consider an arbitrary general dimension CFT 4-point correlator $\mathcal{G}(u,v)$ with a scalar of dimension $\Delta_i$ at the external point $x_i$ (where $i=1,2,3,4$).  We write  its inverse Mellin transform as:
\begin{align}
    \label{appendixMellin:1}
   \mathcal{G}(u,v) =& \int [ds][dt]\, u^s v^t\, 
\Gamma(s+t+a)\,\Gamma(s+t+b)\,\nonumber\\
&\times\Gamma(-t)\,\Gamma(-t-a-b)\,
\Gamma\!\big(\mbox{$\frac{\Delta_1 + \Delta_2}{2} - s$}\big)
\Gamma\!\big(\mbox{$\frac{\Delta_3 + \Delta_4}{2} - s$}\big)
M(s,t),
\end{align}
where, $a = \frac{\Delta_2 - \Delta_1}{2},  b = \frac{\Delta_3 - \Delta_4}{2}$.  Here $[u]$, $[v]$ denote the usual infinite line integral in complex plane parallel to imaginary axis. We refer to $M(s,t)$ as the Mellin amplitude of $\mathcal{G}(u,v)$. 
If we want to restrict to 1d CFT correlators we should set $u=z^2, v=(1-z)^2$. Below we define Witten diagrams in AdS$_2$ in terms of their Mellin amplitudes. We refer the reader to \cite{penedones2011writing, Paulos_2011, Gopakumar:2018xqi} for further details on this topic.

\subsection{Spin 0}
The Mellin amplitude of an AdS$_2$ Witten diagram with an external operator of dimension $\Delta_i$ at $x_i$  which exchanges a spin 0 bulk operator corresponding to dimension $\Delta$ in the s channel ($12\to 34$) is given by
\be\label{Ms-gen}
M^{12,34}_{\Delta,0}(s,t)=N_{\D}^{\{\D_i\}}\sum_{m=0}^\infty \frac{\Gamma \left(\frac{\Delta +\Delta _1+\Delta _2-1}{2} \right) \Gamma \left(\frac{\Delta +\Delta _3+\Delta _4-1}{2} \right) \left(\frac{\Delta -\Delta _1-\Delta _2}{2} +1\right)_m \left(\frac{\Delta -\Delta _3-\Delta _4}{2}+1\right)_m}{2 \Gamma (m+1) \Gamma \left(m+\Delta +\frac{1}{2}\right) (\Delta +2 m-2 s)}\,.
\ee
The normalization (fixed by the coefficient of $G_{\Delta}(z)$ in the block decomposition, see e.g. \eqref{block-deco} below) is given by 
\be
N_{\D}^{\{\D_i\}}=\frac{4 \Gamma (\Delta ) \Gamma \left(\Delta +\frac{1}{2}\right)}{\Gamma \left(\frac{\Delta }{2}\right)^4 \Gamma \left(\frac{\Delta -1}{2}+\Delta _1\right) \Gamma \left(\Delta _1-\frac{\Delta }{2}\right) \Gamma \left(\frac{\Delta -1}{2}+\Delta _2\right) \Gamma \left(\Delta _2-\frac{\Delta }{2}\right)}\,.
\ee
To obtain the diagram that has the same exchange in the t-channel ($14\to 23$) we replace:
\be
M^{14,32}_{\Delta,0}(s,t)=M^{(s)}_{\Delta,0}(s,t)\Big|_{\left\{\Delta _2\leftrightarrow \Delta _4,s\to \frac{\Delta _2+\Delta _3}{2}+t,t\to s-\frac{\Delta _3+\D_4}{2}\right\}}\,,
\ee
and for the u-channel diagram ($13\to 24$) we replace:
\be
M^{13,24}_{\Delta,0}(s,t)=M^{(s)}_{\Delta,0}(s,t)\Big|_{\left\{\Delta_2\leftrightarrow \Delta_3,s\to \frac{\Delta _1+\Delta _4}{2}-s-t,t\to t\right\}}\,.
\ee
\subsubsection{Identical external scalars}
Consider the case $\Delta_i=\Delta_\phi$ that gives the Witten diagrams for identical external scalars. Its conformal block decomposition is obtained by evaluating the residues at all poles of $s$. In position space the decomposition is given by
\be\label{block-deco}
W^{(s)}_{\D,0}(z)=G_\D(z)+\sum_{n=0}^\infty \big(a_n^{(s)}(\D) G_{2\D_\phi+2n}(z)+ b_n^{(s)}(\D) \partial_\Delta G_{2\D_\phi+2n}(z)\big)\,.
\ee
The first term is obtained from the poles $s=\frac{\Delta}{2}+m$ in the Mellin amplitude. The other terms are obtained from the poles in the Mellin measure (see \eqref{appendixMellin:1}) $s=\Delta_\phi+n$. 

The crossed channel diagrams have a block decomposition given by 
\begin{align}
W^{(t)}_{\D,0}(z)=&\sum_{n} \big(a_n^{(t)}(\D) G_{2\D_\phi+2n}(z)+ b_n^{(t)}(\D) \partial_\D G_{2\D_\phi+2n}(z)\big)\nonumber\\
&\sum_{n} \big(\tilde{a}_n^{(t)}(\D) G_{2\D_\phi+2n+1}(z)+ \tilde{b}_n^{(t)}(\D) \partial_\D G_{2\D_\phi+2n+1}(z)\big)\,,
\end{align}
and
\begin{align}
W^{(u)}_{\D,0}(z)=&\sum_{n} \big(a_n^{(u)}(\D) G_{2\D_\phi+2n}(z)+ b_n^{(u)}(\D) \partial_\D G_{2\D_\phi+2n}(z)\big)\nonumber\\
&\sum_{n} \big(\tilde{a}_n^{(u)}(\D) G_{2\D_\phi+2n+1}(z)+ \tilde{b}_n^{(u)}(\D) \partial_\D G_{2\D_\phi+2n+1}(z)\big)\,.
\end{align}
It can be verified that $a_n^{(u)}=a_n^{(t)}$, $b_n^{(u)}=b_n^{(t)}$ whereas $\tilde{a}_n^{(u)}=-\tilde{a}_n^{(t)}$ and $\tilde{b}_n^{(u)}=-\tilde{b}_n^{(t)}$\,. 
The coefficients $a_n^{(s)}$ and $b_{n}^{(s)}$ involve $t$-integrals that can be evaluated readily with the Mellin-Barnes lemma. The coefficients for crossed channel diagrams involve a more nontrivial $t$-integral of the schematic form
\be\label{t-integral-trick}
\int\, [dt] \sum_{m=0}^\infty\frac{\Gamma(X_1+t)\,\Gamma(X_2+t)\,\Gamma(X_3-t)\,\Gamma(X_4-t)}{t+Y+m}X_5(m)\,.
\ee
An efficient way to deal with such integrals is to replace the denominator $(t+Y+m)^{-1}=\int_0^1\, dy\,y^{t+Y+m-1}$\,. Now the $t$-integral can be carried out which gives a ${}_2F_1$ hypergeometric function. The $m$-sum can be carried out independently and gives another  ${}_2F_1$ function. The remaining $y$-integral is simple to evaluate numerically.  \\ \\
We do not show the explicit forms of the block decomposition coefficients as they are straightforward to obtain and numerically evaluate on \textsc{Mathematica} \cite{Mathematica}\,.

\subsubsection{Pairwise identical scalars}\label{pairwisespin0}
We have also used Polyakov blocks for non-identical scalars in the main text. But we have restricted ourselves to the case of two distinct external dimensions: $\Delta_1$ and $\Delta_2$. Consider the case where we have $\Delta_1$ at $x_1, x_2$ and $\Delta_2$ at $x_3, x_4$. This corresponds to replacing in \eqref{Ms-gen}
\be
M^{11,22}_{\D,0}(s,t)=M^{12,34}_{\D,0}(s,t)\Big|_{\{\Delta_1,\Delta_2 \to \Delta_1, \, \Delta_3,\Delta_4 \to \Delta_2\}}\,.
\ee
The crossed channel diagrams are given by
\begin{align}\label{t-pairwiseid}
&M^{12,21}_{\D,0}(s,t)=M^{14,32}_{\D,0}(s,t)\Big|_{\{\Delta_1,\Delta_2 \to \Delta_1, \, \Delta_3,\Delta_4 \to \Delta_2\}}\,,\\
&M^{12,12}_{\D,0}(s,t)=M^{13,24}_{\D,0}(s,t)\Big|_{\{\Delta_1,\Delta_2 \to \Delta_1, \, \Delta_3,\Delta_4 \to \Delta_2\}}\,.
\end{align}
These three diagrams lead to the following s-channel ($11\to 22$) block decompositions in position space
\begin{align}
W^{c}_{\D,0}(z)\, =\, &G_{\D}(z)\, \delta_{c,(11,22)}+\sum_{n} \big(a_n^{c,11}(\D) G_{2\D_1+2n}(z)+ a_n^{c,22}(\D) G_{2\D_2+2n}(z)\big)\nonumber\\ &+\sum_{n} \big(\tilde{a}_n^{c,11}(\D) G_{2\D_1+2n+1}(z)+ \tilde{a}_n^{c,22}(\D) G_{2\D_2+2n+1}(z)\big)
\,.
\end{align}
Here $c=(11,22), (12,21), (12,12)$ denotes the exchange channels. Note that for this configuration of external points the conformal blocks in direct channel are given by $G_{\Delta}(z)=z^{\Delta} \, {}_2F_1(\Delta,\Delta,2\Delta,z)$\,. \\ \\
The decomposition coefficients shown above are obtained by plugging in the respective Mellin amplitude in \eqref{appendixMellin:1} and collecting residues of $s$ variable poles. We point out the relation between expansion coefficients:  $\tilde{a}^{(11,22),11}_n=\tilde{a}^{(11,22),22}_n=0$ and $a^{(12,21),11}_n=a^{(12,12),11}_n$, $a^{(12,21),22}_n=a^{(12,12),22}_n$, $\tilde{a}^{(12,21),11}_n=-\tilde{a}^{(12,12),11}_n$, $\tilde{a}^{(12,21),22}_n=-\tilde{a}^{(12,12),22}_n$.   \,. \\ \\
The diagrams have different t-channel ($12\to 21$) block decompositions, which are given by
\begin{align}
W^c_{\D,0}(z)=&G^{(t)}_{\D}(z)\, \delta_{c,(12,21)}+\sum_{n} \big(a_n^{c}(\D) G^{(t)}_{\D_1+\D_2+2n}(z)+ b_n^{c}(\D) \partial_\D G^{(t)}_{\D_1+\D_2+2n}(z)\big)\nonumber\\
&\sum_{n} \big(\tilde{a}_n^{c}(\D) G^{(t)}_{\D_1+\D_2+2n+1}(z)+ \tilde{b}_n^{c}(\D) \partial_\D G^{(t)}_{\D_1+\D_2+2n+1}(z)\big)\,.
\end{align}
The t-channel blocks are given by $G^{(t)}_{\Delta}(z)=z^{\Delta-\Delta_1-\Delta_2} \, {}_2F_1(\Delta-\Delta_{12},\Delta-\Delta_{12},2\Delta,1-z)$\,. The corresponding decomposition coefficients are obtained by collecting residues of $t$ variable poles in \eqref{appendixMellin:1}. \\ \\
For the pairwise identical case  whenever the exchange channel and the decomposition channel are different, there is a Mellin integral of the form \eqref{t-integral-trick}. They may be evaluated by the same trick discussed below that equation. 

\subsubsection{Orthogonality}
The coefficients appearing in the pairwise identical case have orthogonality, meaning they vanish at locations of specific double twist operators. The direct channel exhange Witten diagram, one finds (we denote $\Delta_i=\Delta_1,\Delta_2$):
\begin{align}
&a_n^{(11,22),11}(2\Delta_1+2m)\propto \delta_{mn}\,, &\quad&a_n^{(11,22),11}(2\Delta_2+2m)=0\,,\nonumber\\
&a_n^{(11,22),22}(2\Delta_1+2m)=0\,, &\quad&a_n^{(11,22),22}(2\Delta_2+2m)\propto \delta_{mn}\,,\nonumber\\
&a_n^{(11,22)}(2\Delta_i+2m)=0\,, &\quad&b_n^{(11,22)}(2\Delta_i+2m)=0\,,\nonumber\\
&\tilde{a}_n^{(11,22)}(2\Delta_i+2m+1)=0\,, &\quad&\tilde{b}_n^{(11,22)}(2\Delta_i+2m+1)=0\,.
\end{align}
On the other hand the orthogonality conditions in the crossed channel diagrams are of the form
\begin{align}\label{1221exchangeortho}
&a_n^{(12,21),11}(\Delta_1+\Delta_2+2m)=0\,, &\quad&\partial_\Delta a_n^{(12,21),11}(\Delta_1+\Delta_2+2m)=0\,,\nonumber\\
&a_n^{(12,21),22}(\Delta_1+\Delta_2+2m)=0\,, &\quad&\partial_\Delta a_n^{(12,21),22}(\Delta_1+\Delta_2+2m)=0\,,\nonumber\\
&a_n^{(12,21)}(\Delta_1+\Delta_2+2m)\propto \delta_{mn}\,, &\quad&\partial_\Delta a_n^{(12,21)}(\Delta_1+\Delta_2+2m)=0\,,\nonumber\\
&b_n^{(12,21)}(\Delta_1+\Delta_2+2m)=0\,, &\quad&\partial_\Delta b_n^{(12,21)}(\Delta_1+\Delta_2+2m)\propto\delta_{mn}\,,\nonumber\\
&\tilde{a}_n^{(12,21)}(\Delta_1+\Delta_2+2m+1)\propto \delta_{mn}\,, &\quad&\partial_\Delta \tilde{a}_n^{(12,21)}(\Delta_1+\Delta_2+2m+1)=0\,,\nonumber\\
&\tilde{b}_n^{(12,21)}(\Delta_1+\Delta_2+2m+1)=0\,, &\quad&\partial_\Delta \tilde{b}_n^{(12,21)}(\Delta_1+\Delta_2+2m+1)\propto\delta_{mn}\,.
\end{align}
The orthogonality of the $(12,12)$ channel exchange is identical to \eqref{1221exchangeortho} except that one always has 0 on the r.h.s.. 
Notice that the orthogonality of a $c$-channel exchange diagram is always with respect to the double twist operators relevant to the $c$ channel and irrespective of those of the channel of decomposition.

\subsection{Spin 1 exchange}
The Mellin amplitude of a  Witten diagram, with external operators of dimension $\Delta_i$ at  $x_i$,  which exchanges a spin $1$ bulk operator corresponding to dimension $\Delta$ in the s channel $(12 \rightarrow 34)$ is given by

\begin{equation}
\begin{aligned}
    \label{Ms-spin1}
    M_{\Delta,1}^{12,34}(s,t)= N_{\Delta}^{\{\Delta_i\}} \sum_{n=0}^\infty \frac{\Gamma \left(\frac{\Delta+\Delta_1+\Delta_2}{2}\right) \Gamma \left(\frac{\Delta+\Delta_3+\Delta_4}{2}\right) \left(\frac{\Delta-\Delta_1-\Delta_2+1}{2}\right)_n \left(\frac{\Delta-\Delta_3-\Delta_4+1}{2}\right)_n(pt +q+s)}{(\Delta -1)\Delta \Gamma(n+1)\left(\Delta + \frac{1}{2}\right)_n (\Delta + 2n -2s -1)},
\end{aligned}
\end{equation}
where
\begin{equation}
\begin{aligned}
    \label{p-def}
    p=&\frac{2 (\Delta -1) \Delta }{(\Delta -1) \Delta -\left(\Delta _1-\Delta _2\right) \left(\Delta _3-\Delta _4\right)},\\
    q=& \frac{\left(-\Delta _1+\Delta _2+\Delta _3-\Delta _4\right) \Delta ^2+\left(\Delta _1-\Delta _2-\Delta _3+\Delta _4\right) \Delta -\left(\Delta _1-\Delta _2\right) \left(\Delta _3-\Delta _4\right)}{2 (\Delta -1) \Delta -2 \left(\Delta _1-\Delta _2\right) \left(\Delta _3-\Delta _4\right)},
\end{aligned}
\end{equation}
and here the normalisation 

\begin{equation}
\begin{aligned}
    \label{sp1normalisation}
    N_{\Delta}^{\{\Delta_i\}}=\frac{\left(2\Gamma \left(\frac{\Delta +\Delta _1-\Delta _2+1}{2} \right) \Gamma \left(\frac{\Delta -\Delta _1+\Delta _2+1}{2}\right)\right)^{-1}(\Delta -1) \left((\Delta -1) \Delta -\left(\Delta _1-\Delta _2\right) \left(\Delta _3-\Delta _4\right)\right) \Gamma (\Delta +1)}{  \Gamma \left(\frac{\Delta _1+\Delta _2+1-\Delta }{2}\right) \Gamma \left(\frac{\Delta +\Delta _1+\Delta _2}{2} \right) \Gamma \left(\frac{\Delta +\Delta _3-\Delta _4+1}{2} \right) \Gamma \left(\frac{\Delta -\Delta _3+\Delta _4+1}{2} \right) \Gamma \left(\frac{\Delta _3+\Delta _4+1-\Delta}{2} \right) \Gamma \left(\frac{\Delta +\Delta _3+\Delta _4}{2}\right)}.
\end{aligned}
\end{equation}
Similar to the spin 0 case, the same spin $1$ exchange diagram in t channel $(14 \rightarrow 23)$ is given by
\begin{equation}
\begin{aligned}
    \label{sp1M1234t}
    M^{14,32}_{\Delta,1}(s,t)=M^{(s)}_{\Delta,1}(s,t)\Big|_{\left\{\Delta _2\leftrightarrow \Delta _4,s\to \frac{\Delta _2+\Delta _3}{2}+t,t\to s-\frac{\Delta _3+\Delta_4}{2}\right\}}
\end{aligned}
\end{equation}
and the one in u channel $(13 \rightarrow 24)$ by
\begin{equation}
\begin{aligned}
    \label{sp1M12234u}
    M^{13,24}_{\Delta,1}(s,t)=M^{(s)}_{\Delta,1}(s,t)\Big|_{\left\{\Delta_2\leftrightarrow \Delta_3,s\to \frac{\Delta _1+\Delta _4}{2}-s-t,t\to t\right\}}
\end{aligned}
\end{equation}
Now, if we make them pairwise identical, it looks simpler. This corresponds to replacing in \eqref{Ms-spin1}
\begin{equation}
\begin{aligned}
    \label{sp1M1122}
    M^{11,22}_{\Delta,1}(s,t)=M^{12,34}_{\Delta,1}(s,t)\Big|_{\{\Delta_1,\Delta_2 \to \Delta_1, \, \Delta_3,\Delta_4 \to \Delta_2\}}.
\end{aligned}
\end{equation}
The crossed channel diagrams follow by the same replacement (compare with \eqref{t-pairwiseid}). 
Having all these three diagrams, we have the s-channel block decomposition as follows
\begin{equation}
\begin{aligned}
    \label{sp1Witts}
    W^{c}_{\Delta,1}(z)\, =\, &G_{\Delta}(z)\, \delta_{c,(11,22)}+\sum_{n} \big(A_n^{c,11}(\Delta) G_{2\Delta_1+2n+1}(z)+ A_n^{c,22}(\Delta) G_{2\Delta_2+2n+1}(z)\big)\\&+\sum_{n} \big(\tilde{A}_n^{c,11}(\Delta) G_{2\Delta_1+2n}(z)+ \tilde{A}_n^{c,22}(\Delta) G_{2\Delta_2+2n}(z)\big)\,. 
\end{aligned}
\end{equation}
Here, $c= (11,22), (12,21) \text{ or } (12,12)$ denotes the exchange channels of the diagrams. \\ \\ 
As in the spin 0 case, there are special relations in the decomposition coefficients: $A^{(11,22),11}_n=A^{(11,22),22}_n=0$ and $A^{(12,21),11}_n=-A^{(12,12),11}_n$, $A^{(12,21),22}_n=-A^{(12,12),22}_n$, $\tilde{A}^{(12,21),11}_n=\tilde{A}^{(12,12),11}_n$, $\tilde{A}^{(12,21),22}_n=\tilde{A}^{(12,12),22}_n$.   \,.  \\ \\
The diagrams decompose into t-channel blocks in the following way
\begin{equation}
\begin{aligned}
    \label{sp1WittT}
    W_{\Delta,1}^c(z)=G_{\Delta}^{(t)}\delta_{c,(12,21)}+&\sum_{n} \big(A_n^{c}(\Delta) G^{(t)}_{\Delta_1+\Delta_2+2n+1}(z)+ B_n^{c}(\Delta) G^{(t)}_{\Delta_1+\Delta_2+2n+1}(z)\big)\\
& \sum_{n} \big(\tilde{A}_n^{c}(\Delta) G^{(t)}_{\Delta_1+\Delta_2+2n}(z)+ \tilde{B}_n^{c}(\Delta) G^{(t)}_{\Delta_1+\Delta_2+2n}(z)\big)\,.
\end{aligned}
\end{equation}
One still has orthogonality properties but now they are at slightly different locations. For the direct channel exchange diagram one has 
\begin{align}\label{1122exchangeorthoA}
&A_n^{(11,22),11}(2\Delta_1+2m+1)\propto \delta_{mn}\,, &\quad&A_n^{(11,22),11}(2\Delta_2+2m+1)=0\,,\nonumber\\
&A_n^{(11,22),22}(2\Delta_1+2m+1)=0\,, &\quad&A_n^{(11,22),22}(2\Delta_2+2m+1)\propto \delta_{mn}\,,\nonumber\\
&A_n^{(11,22)}(2\Delta_i+2m+1)=0\,, &\quad&B_n^{(11,22)}(2\Delta_i+2m+1)=0\,,\nonumber\\
&\tilde{A}_n^{(11,22)}(2\Delta_i+2m)=0\,, &\quad&\tilde{B}_n^{(11,22)}(2\Delta_i+2m)=0\,,
\end{align} 
while for $12,21$ channel exchange one has
\begin{align}\label{1221exchangeorthoA}
&A_n^{(12,21),11}(\Delta_1+\Delta_2+2m+1)=0\,, &\quad&\partial_\Delta A_n^{(12,21),11}(\Delta_1+\Delta_2+2m+1)=0\,,\nonumber\\
&A_n^{(12,21),22}(\Delta_1+\Delta_2+2m+1)=0\,, &\quad&\partial_\Delta A_n^{(12,21),22}(\Delta_1+\Delta_2+2m+1)=0\,,\nonumber\\
&A_n^{(12,21)}(\Delta_1+\Delta_2+2m+1)\propto \delta_{mn}\,, &\quad&\partial_\Delta A_n^{(12,21)}(\Delta_1+\Delta_2+2m+1)=0\,,\nonumber\\
&B_n^{(12,21)}(\Delta_1+\Delta_2+2m+1)=0\,, &\quad&\partial_\Delta B_n^{(12,21)}(\Delta_1+\Delta_2+2m+1)\propto\delta_{mn}\,,\nonumber\\
&\tilde{A}_n^{(12,21)}(\Delta_1+\Delta_2+2m)\propto \delta_{mn}\,, &\quad&\partial_\Delta \tilde{A}_n^{(12,21)}(\Delta_1+\Delta_2+2m)=0\,,\nonumber\\
&\tilde{B}_n^{(12,21)}(\Delta_1+\Delta_2+2m)=0\,, &\quad&\partial_\Delta \tilde{B}_n^{(12,21)}(\Delta_1+\Delta_2+2m)\propto\delta_{mn}\,.
\end{align}
Here too the $12,12$ channel exchange orthogonality are identical to \eqref{1221exchangeorthoA} except that $\delta_{mn}$ is replaced by 0 on the r.h.s. everywhere. 


\subsection{Contact diagrams}
Let us now look at the contact Witten diagrams.
The most generic Mellin amplitude that can be written for a contact Witten diagram preserving Regge boundedness \eqref{Regge} with all non-identical external legs, is
\begin{equation}
    \label{contactMellin}
    M(s,t)= a_1(\Delta_1,\Delta_2,\Delta_3,\Delta_4)s+a_2(\Delta_1,\Delta_2,\Delta_3,\Delta_4)t+ a_3(\Delta_1,\Delta_2,\Delta_3,\Delta_4)\,.
\end{equation}
These coefficients $a_1, a_2, a_3$ are independent but should be respecting crossing symmetry. These three constants correspond to three independent coupling constants for quartic vertices - one with zero derivatives and two with two derivatives.\\ \\
For the identical scalar case ($\Delta_i\to \Delta_\phi$) the Mellin amplitude of a contact diagram is just a constant. We can take $M^{\text{(con)}}(s,t)=1$. The corresponding diagram in position space decomposes into 
\be\label{contact-block-deco}
W^{\text{(con)}}(z)=\sum_{n=0}^\infty \big(a_n^{(\text{con})} G_{2\D_\phi+2n}(z)+ b_n^{(\text{con})} \partial_\Delta G_{2\D_\phi+2n}(z)\big)\,.
\ee
If we similarly focus on the pairwise identical case $(\Delta_1,\Delta_2 \rightarrow \Delta_1, \Delta_3, \Delta_4 \rightarrow \Delta_2)$, we get a simple Mellin amplitude given by
\begin{equation}
    \label{contact1122}
    M^{\text{(con)}}(s,t)= a s+b\,.
\end{equation}
There is no $t$ in this Mellin amplitude as it preserves the t $\leftrightarrow$ u channel symmetry which corresponds to symmetry under replacing $t\leftrightarrow s+t$. \\ \\
In position space the corresponding Witten diagram has an s-channel decomposition
\begin{equation}
\begin{aligned}
    \label{contact}
     W^{(\text{con})}(z)=\sum_{n=0}^\infty \big(c_n^{(c,11)}\, G_{2\Delta_1+2n}(z)+ c_n^{(c,22)}\, G_{2\Delta_2+2n}(z)\big)\,.
\end{aligned}
\end{equation}
The same diagram has a t-channel decomposition 
\begin{equation}
\begin{aligned}
    \label{contact1221}
    W^{(\text{con})}(z)=\sum_{n=0}^\infty \big(c_n \, G^{(t)}_{\Delta_1+\Delta_2+2n}(z)+ d_n\, \partial _{\Delta} G^{(t)}_{\Delta_1+\Delta_2+2n}(z)\big)\\
    +\sum_{n=0}^\infty \big(\tilde{c}_nG^{(t)}_{\Delta_1+\Delta_2+2n+1}(z)+ \tilde{d}_n\, \partial_{\Delta}G^{(t)}_{\Delta_1+\Delta_2+2n+1}(z)\big) \,. 
\end{aligned}
\end{equation}

\section{Cubic Hermite Interpolation}
\label{sec:CHIP}
Interpolation refers to the broad class of methods for determining a function
\begin{equation}
\mathbf{f}\quad :\quad  \mathcal{X} \longrightarrow
\mathcal{Y}\,,
\quad \mathcal{X}\subset \mathbb{R}\,,\,
\mathcal{Y} \subset \mathbb{R}\,,
\end{equation}
from \(N\) observations or \emph{nodes}
\begin{equation}
\left(x_i,y_i\right)\in\mathcal{X}
\times \mathcal{Y}\,,\,\qquad 
i=1,2,\,\ldots, N
\,.
\end{equation}
We assume the nodes to be ordered, in the sense that \(x_{i+1}>x_i\) for all \(i\)\,. In the following, we consider a slight extension of the problem where the available data at each node is a triplet
\begin{equation}
    \left(x_i,\,y_i,\,m_i\right)\,,
\end{equation}
where \(m\) is the derivative of the interpolation polynomial. That is, the goal is to determine that 
\begin{equation}
\mathbf{f}\left(x_i\right) = y_i\,,\quad
\mathbf{f}^\prime\left(x_i\right) = m_i\,.
\end{equation}
Cubic Hermite Interpolation solves this problem in the following way. Consider the real interval \(t\in\left[0,1\right]\) and define the interpolation data across the end points
\begin{equation}
\{\left(0,p_0,m_0\right)\,,\,
\left(1,p_1,m_1\right)\}\,.
\end{equation}
An interpolating polynomial is 
\begin{equation}\label{eq:CHIP}
\mathrm{p}\left(t\right) = 
\mathrm{h}_{00}\left(t\right)\mathrm{p}_0
+\mathrm{h}_{01}\left(t\right)\mathrm{p}_1
+\mathrm{h}_{10}\left(t\right)\mathrm{m}_0
+\mathrm{h}_{11}\left(t\right)\mathrm{m}_1\,,
\end{equation}
where
\begin{equation}
\begin{split}
\mathrm{h}_{00}\left(t\right)
= 2t^3-3t^2+1\,,
\quad& \mathrm{h}_{01}\left(t\right)
=-2t^3+3t^2\,,\\
\mathrm{h}_{10}\left(t\right)
= t^3-2t^2+t\,,
\quad &
\mathrm{h}_{11}\left(t\right)
= t^3-t^2\,,
\end{split}
\end{equation}
and hence the \(t\) derivatives are
\begin{equation}
\mathrm{h}^\prime_{00}
= 6\,t\left(t-1\right)\,,\quad
\mathrm{h}^\prime_{01}
=- 6\,t\left(t-1\right)\,,\quad
\mathrm{h}^\prime_{10}
= 3t^2-4t+1\,,\quad
\mathrm{h}^\prime_{11}
= t\left(3t-2\right)\,.
\end{equation}
The basis functions 
\(h_{ij}\) obey the boundary values
\begin{equation}
\left(h_{00}, h_{01},h_{10},h_{11}\right)
=\left(1,\,1,\,0,0\right)\,,\quad 
\left(h^\prime_{00}, h^\prime_{01},h^\prime_{10},h^\prime_{11}\right)
=\left(0,\,0,\,1,1\right)\,,
\end{equation}
and hence the boundary conditions arising from the interpolation data are satisfied. This extends to the interpolation problem over the interval \(\mathcal{I}_k\) by identifying \(t\in\left[0,1\right]\) to
\begin{equation}
t = \frac{x-x_k}{x_{k+1}-x_k}\,,
\end{equation}
so that \(t=0\) corresponds to \(x=x_k\) 
and \(t=1\) corresponds to \(x=x_{k+1}\)\,. 
The polynomial \eqref{eq:CHIP} can also be written as
\begin{equation}
\mathrm{p}\left(t\right) = 
\left(2\,\mathrm{p}_0-2\mathrm{p}_1
+\mathrm{m}_0+\mathrm{m}_1\right)t^3
-\left(3\mathrm{p}_0-3\mathrm{p}_1+2\mathrm{m}_0+\mathrm{m}_1\right)t^2
+\mathrm{m}_0t+\mathrm{p}\,.
\end{equation}
In this way, a piecewise cubic interpolant 
through the given nodes can be constructed. 
The formula is
\begin{equation}
\mathrm{p}\left(x\right) = 
\mathrm{h}_{00}\left(t\right)\mathrm{p}_k
+\mathrm{h}_{01}\left(t\right)
\mathrm{p}_{k+1} +
\delta x_{k}\,
\mathrm{h}_{10}\left(t\right)\mathrm{m}_k
+\delta x_{k}\,\mathrm{h}_{11}
\left(t\right)\mathrm{m}_{k+1}\,.
\end{equation}

\section{Optimization by Gradient Descent}

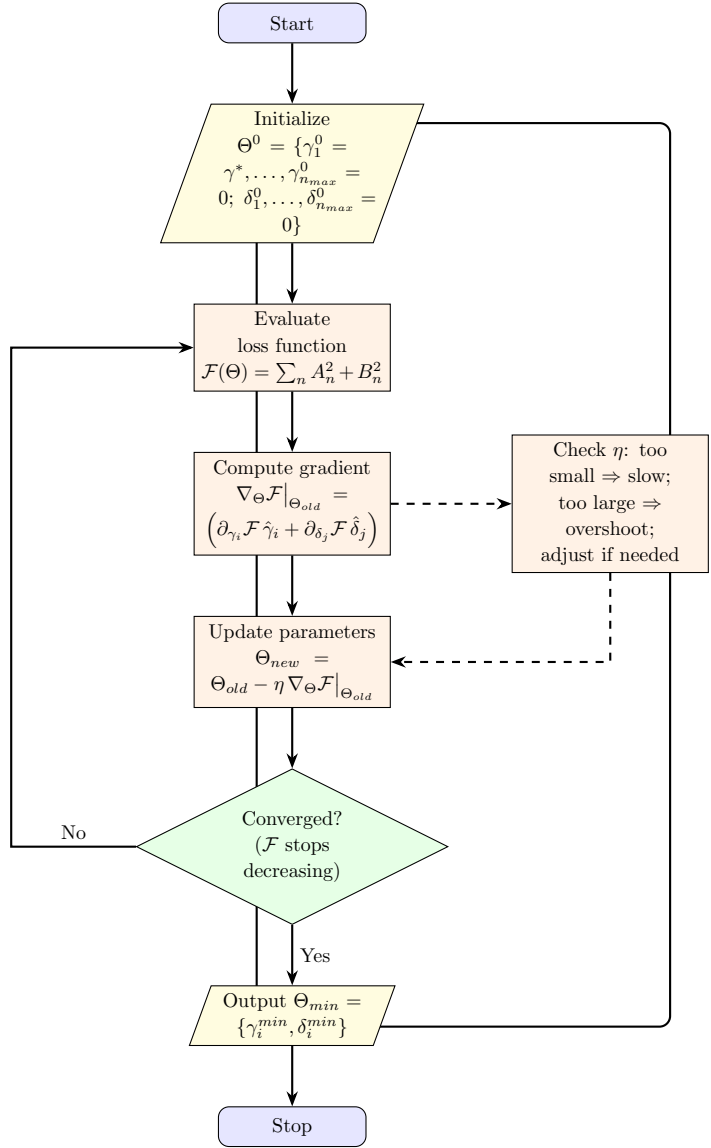
\begin{figure}[h]
\centering
\begin{tikzpicture}[
    scale=0.75, every node/.style={scale=0.75},
    node distance = 0.8cm and 1.3cm,
    startstop/.style = {rectangle, rounded corners, minimum width=2.6cm, minimum height=0.7cm,
                         text centered, draw=black, fill=blue!10, font=\small},
    process/.style   = {rectangle, minimum width=3.2cm, minimum height=0.8cm,
                         text centered, text width=3.2cm, draw=black, fill=orange!10, font=\small},
    decision/.style  = {diamond, aspect=2, minimum width=2.6cm, minimum height=1cm,
                         text centered, text width=2.2cm, draw=black, fill=green!10, font=\small},
    io/.style        = {trapezium, trapezium left angle=70, trapezium right angle=110,
                         minimum width=2.6cm, minimum height=0.7cm, text centered,
                         text width=2.6cm, draw=black, fill=yellow!15, font=\small},
    arrow/.style      = {-{Stealth[length=2mm]}, thick}
]

\node (start) [startstop] {Start};

\node (init) [io, below=of start]
    {Initialize $\Theta^0 = \{\gamma_1^0=\gamma^*,\dots,\gamma_{n_{max}}^0=0;\ \delta_1^0,\dots,\delta_{n_{max}}^0=0\}$};

\node (evalF) [process, below=of init]
    {Evaluate loss function \\ $\mathcal{F}(\Theta) = \sum_{n} A_n^2 + B_n^2$};

\node (grad) [process, below=of evalF]
    {Compute gradient \\ $\nabla_{\Theta}\mathcal{F}\big|_{\Theta_{old}}
     = \left(\partial_{\gamma_i}\mathcal{F}\,\hat{\gamma}_i + \partial_{\delta_j}\mathcal{F}\,\hat{\delta}_j\right)$};

\node (update) [process, below=of grad]
    {Update parameters \\ $\Theta_{new} = \Theta_{old} - \eta\, \nabla_{\Theta}\mathcal{F}\big|_{\Theta_{old}}$};

\node (check) [decision, below=of update]
    {Converged? ($\mathcal{F}$ stops decreasing)};

\node (etacheck) [process, right=1.6cm of grad]
    {Check $\eta$: too small $\Rightarrow$ slow; \\
     too large $\Rightarrow$ overshoot; \\ adjust if needed};

\node (output) [io, below=of check]
    {Output $\Theta_{min} = \{\gamma_i^{min}, \delta_i^{min}\}$};

\node (stop) [startstop, below=of output] {Stop};

\draw [arrow] (start) -- (init);
\draw [arrow] (init) -- (evalF);
\draw [arrow] (evalF) -- (grad);
\draw [arrow] (grad) -- (update);
\draw [arrow] (update) -- (check);
\draw [arrow] (check) -- node[anchor=west, font=\small]{Yes} (output);
\draw [arrow] (output) -- (stop);

\draw [arrow] (check.west) -- ++(-2.2,0) node[midway, above, font=\small]{No} |- (evalF.west);

\draw [arrow, dashed] (grad) -- (etacheck);
\draw [arrow, dashed] (etacheck) |- (update);

\begin{scope}[on background layer]
\node [draw, thick, rounded corners, inner sep=0.4cm,
       fit=(start)(stop)(etacheck)] {};
\end{scope}

\end{tikzpicture}
\caption{Gradient-descent optimization flow used in the truncated bootstrap.}
\label{fig:gd-flowchart}
\end{figure}

In this section, we briefly try to give an idea of how the ``Gradient-Descent" algorithm actually works.
It is an optimization algorithm used to minimize a loss function or target function.
Let's say we have a loss function $\mathcal{F}(\Theta)$ , where $\Theta = \{\gamma_i,\delta_j\}$, which we want to minimize.
Starting with some initial values, $\Theta^0=\{\gamma_i^0,\delta_j^0\}$ we have,
\begin{equation}
    \label{Loss-Func GD}
    \mathcal{F}(\Theta^0)=\mathcal{F}(\{\gamma_1^0,\gamma_2^0,...,\gamma_m^0;\delta_1^0,...,\delta_n^0\}) = F_{in}
\end{equation}
In the parameter space, we now search for the steepest slope of the loss function $\mathcal{F}$.
We calculate,
\begin{equation}
    \label{nabla}
    \nabla_{\Theta}\mathcal{F}(\gamma_i,\delta_j)=\left(\partial_{\gamma_i}\mathcal{F}(\gamma_i,\delta_j)\:\hat{\gamma_i}+ \partial_{\delta_j}\mathcal{F}(\gamma_i,\delta_j)\:\hat{\delta_j}\right)\Big|_{\Theta= \Theta_0}
\end{equation}
which is nothing but the gradient of the loss function.
Then we modify the parameter values as below,

\begin{equation}
    \label{grad_des step}
    \Theta_{new} = \Theta_{old}- \eta . \nabla_{\Theta}\mathcal{F}(\Theta)|_{\Theta=\Theta_{old}} 
\end{equation}

Breaking this down:
\begin{itemize}
    \item $\nabla_{\Theta}\mathcal{F}:$ the gradient, a vector of partial derivatives, pointing in the steepest increase in the parameter space.
    \item $\eta:$ The learning rate is a small number defining the step size of the gradient search.
    \item The minus sign: As we want to find the minimum, we go in the opposite direction of the gradient.
\end{itemize}
We repeat this step iteratively until the loss function stops decreasing meaningfully.
This learning rate massively controls this convergence.
If it's too small, the convergence is painfully slow.
If it's too large, we might overshoot the minimum and bounce back and forth.
So we need to set a good $``\eta"$ to get a stable search.

We would like to optimize a target function which arises in our truncated bootstrap as mentioned in \eqref{F-loss}.
It roughly looks like,
\begin{equation}
\mathcal{L}=\sum_{n=1}^{n_{max}} A_n^2+B_n^2\,,
\end{equation}
where \(A_n\) and \(B_n\) are polynomials in \(\delta_i\) and \(\gamma_i\). 
\eqref{F-loss} shows there are two types of functionals - $\alpha_n$ and $\beta_n$. 
According to this convention, $B_n$ corresponds to $\beta_n$  and $A_n$ corresponds to $\alpha_n$  of the loss function $\mathcal{F}$.
The overall structure of the functions is
\begin{equation}
\begin{aligned}
     \label{A_n B_n}
     B_n &=\sum_{i=1}^{n_{max}}\left\lbrace \sum_{j=0}^{30}t^{(i)}_{n}\left(2\Delta_{\phi}+2i+\gamma_i\right)^j\right\rbrace \left(c_i+\delta_i\right)\\
     A_n &=\sum_{i=2}^{n_{max}}\left\lbrace \sum_{j=0}^{30}q^{(i)}_{n}\left(2\Delta_{\phi}+2i+\gamma_i\right)^j\right\rbrace \left(c_i+\delta_i\right)
\end{aligned}
\end{equation}

We chose to set $\gamma_1$ as a regulator of our problem which we fix by hand and leave the rest of the parameters as the free parameters of the optimization problem.
So, we have in total $n_{max}$ no. of $\alpha_n[\Delta]$ functionals and $(n_{max}-1)$ no. of $\beta_n[\Delta]$ functionals inside the loss function $\mathcal{F}$.
The index $``j"$ runs over the order of polynomial$($ in our case Hermite polynomials$)$ we choose to fit for the functionals and $``i"$ runs over the number of functionals we work with.

We start with some parameter values $\Theta^0 = \{\gamma_1^0= \gamma^* ,...,\gamma^0_{n_{max}}= 0, \delta_1^0=0,...,\delta^0_{n_{max}}=0\}$.
If every single parameter was identically equal to zero, it would have been trivially sitting on a minima because it would correspond to the generalized free field theory.
Having $\gamma_1^0 \neq0$, we search for a local minima of $\mathcal{F}$ around these parameter values using the element $\nabla_{\Theta}\mathcal{F}$ from \eqref{grad_des step}.
The construction of the loss function \eqref{F-loss} shows that we want to simultaneously minimize all $(2n_{max}-1)$ sum rules.
Squaring $A_n$ and $B_n$ and searching for their minima automatically minimizes each of the sum rules.
The algorithm is shown diagrammatically in Fig. \ref{fig:gd-flowchart}.

Physically these $\gamma_i$ would mean a correction to the anomalous dimensions of the free theory and $\delta_i$ would be the corrections to the OPE coefficients.
Finally we will end up on a CFT data,

\begin{equation}
\begin{aligned}
    \label{optimesed parameters}
    \Delta^{i} &= \Delta_{GFF}^i + \gamma^i = 2\Delta_{\phi}+2n+l+\gamma^i;\\ 
    a_{\Delta_i} &= a_{\Delta_{GFF}}+ \delta_i;
\end{aligned}
\end{equation}

\bibliographystyle{unsrt}
\bibliography{references}

\end{document}